\documentclass{elsart}

\usepackage{graphicx}
\usepackage{amsmath}
\usepackage{bm}  
\usepackage[titles]{tocloft}

\journal{Annals of Physics}

\newcommand{\beq}{\begin{eqnarray}}
\newcommand{\eeq}{\end{eqnarray}}
\newcommand{\bqa}{\begin{eqnarray}}
\newcommand{\eqa}{\end{eqnarray}}

\def\mqo2{{\!\!\!}}

\begin{document}
\begin{frontmatter}

\title{Scattering Models for Ultracold Atoms}

\author[OSU]{Eric Braaten}\ead{braaten@mps.ohio-state.edu}, 
\author[Arizona]{Masaoki Kusunoki}\ead{masa@mps.ohio-state.edu},
\author[OSU]{Dongqing Zhang}\ead{zhangdq@mps.ohio-state.edu}

\address[OSU]{Department of Physics, The Ohio State University,
                Columbus, OH 43210, USA}
\address[Arizona]{Department of Physics,
        University of Arizona,
        Tucson, AZ\ 85721, USA }

\date{\today}

\begin{abstract}
We present a review of scattering models
that can be used to describe the low-energy behavior of
identical bosonic atoms.
In the simplest models, the only degrees of freedom are atoms 
in the same spin state.  More elaborate models 
have other degrees of freedom, such as atoms in other 
spin states or diatomic molecules.
The parameters of the scattering models are specified by giving 
the S-wave phase shifts for scattering
of atoms in the spin state of primary interest.
The models are formulated as local quantum field theories
and the renormalization of their coupling constants is determined.
Some of the parameters can be constrained by renormalizability 
or by the absence of negative-norm states.
The Green's functions that describe the evolution of 2-atom states
are determined analytically.  They are used to determine the 
T-matrix elements for atom-atom scattering and the binding energies 
of diatomic molecules.
The scattering models all exhibit universal behavior as the 
scattering length in a specific spin state becomes large.
\end{abstract}


\end{frontmatter}


\section{Introduction}
\label{sec:intro}

The development of methods for trapping atoms and cooling them 
to ultralow temperatures has launched the new field of 
{\it cold atom physics}.
Among the dramatic achievements in this field have been the 
creation of {\it Bose-Einstein condensates} of bosonic atoms 
and {\it superfluids} of fermionic atoms.
At the low temperatures at which these phenomena occur, 
the de Broglie wavelengths of the atoms are much larger 
than their sizes or the ranges
of their interactions. This makes the behavior of the atoms insensitive
to many of the details of atomic structure and interatomic interactions.
The atoms can be treated as point particles and their interactions can be
described very accurately by simple models.

One class of models that is particularly convenient for describing 
ultracold atoms is {\it scattering models}. 
Scattering models are formulated in
terms of parameters that describe the low-energy scattering of atoms. 
The simplest example is the {\it Zero-Range Model} whose only 
interaction parameter in the 2-body sector is the 
{\it S-wave scattering length}.  The scattering length is the
most important interaction parameter governing the behavior of cold atoms.
The primary advantage of scattering models for describing cold atoms is
that their parameters are directly related
to physical observables at the relevant energy scale.

Another advantage of scattering models is that there are many useful
models that can be solved analytically in the 2-body sector. There are
very few analytic results on the 3-body and higher $n$-body problem, 
but having an analytic solution to the 2-body problem is a great
simplification in the numerical solution of higher few-body problems.
The analytic solution of the 2-body problem is also useful in the
many-body problem, because interaction effects in low-density 
systems of atoms are dominated by 2-body interactions. 

A particularly convenient class of scattering models is 
ones with {\it local interactions}. 
In these models, atoms interact with each other only when they
are at the same point in space. Such a simplified representation of the
interactions for cold atoms is useful because the large de Broglie
wavelengths of the atoms makes them insensitive to the range of
interatomic interactions.  As a consequence, the low-energy behavior of
the atoms can be described to a good approximation by zero-range
interactions.

 A scattering model with local interactions can be described by a 
{\it local quantum field theory}. A formulation in terms of a quantum field
theory is useful because the same framework describes the 2-body, higher
few-body, and many-body problems. Quantum field theories with local
interaction terms are particularly convenient, because theoretical
methods for dealing with such theories are well developed. Much of the
stimulus for this development has come from the success of local
relativistic quantum field theories in describing elementary particles.
One of the complications of local quantum field theories is that they are
inherently singular at short distances, but this problem can be handled
using the machinery of {\it renormalization}.

In this review, we present a unified treatment of scattering models 
that can be used to describe the low-energy behavior of
atoms or other nonrelativistic particles with
short-range interactions. We restrict our attention for the most part to
bosons with the same mass and with S-wave interactions only. The
extensions to fermions, to particles with different masses, and to
interactions in higher angular momentum channels are straightforward
using the formalism of quantum field theory. In the simplest scattering
models, the only degrees of freedom are atoms that are all in the same
spin state. We also consider scattering models whose degrees of freedom
include atoms in other spin states or diatomic molecules.

We define the physical parameters of each scattering model by specifying
the S-wave phase shift for the spin state of primary interest. We
formulate the model as a local quantum field theory with an ultraviolet
cutoff. The renormalizations that relate the coupling constants of the 
quantum field theory to the physical parameters of the scattering model
are determined.  The Green's function that 
describes the evolution of a 2-atom system
is derived analytically.  It is used to determine the
T-matrix elements for atom-atom scattering and the binding energies 
of diatomic molecules. 

We begin the review in Sec.~\ref{sec:fun}
by describing a {\it fundamental theory} that
provides an extremely accurate description of the low-energy behavior of
alkali atoms. We then give a general discussion of 
scattering models in Sec.~\ref{sec:scatmod}.
In Sec.~\ref{sec:zrm}, we discuss the Zero-Range Model,
whose only interaction parameter in the 2-atom sector is the scattering
length. In Sec.~\ref{sec:erm}, we discuss the {\it Effective Range
Model}, which has a second interaction parameter. 
Other scattering models whose
only degrees of freedom are an atom in a single spin state
are described briefly in Sec.~\ref{sec:oscm}. In
Sec.~\ref{sec:2cm}, we discuss the {\it Two-Channel Model}, which
describes atoms that can be in either of two spin states. In
Sec.~\ref{sec:rm}, we describe the {\it Resonance Model}, in which a
diatomic molecule enters as an additional degree of freedom.

\section{The Fundamental Theory}
\label{sec:fun}

In this section, we describe the {\it fundamental theory} that
provides an extremely accurate description of atoms with energies
small compared to the splitting between the ground state
of the atom and its first electronic excitation.
To be specific, we focus on alkali atoms.
These atoms have been used extensively in
cold atom experiments, because they have properties that make
it particularly easy to cool them to ultralow temperatures
using current technology, such as laser cooling and
evaporative cooling.

\subsection{Hamiltonian}
\label{sec:Hfun}

The alkali atoms H, Li, Na, K, Rb, Cs, and Fr can be labeled 
by an integer $n=1$, 2, 3, 4, 5, 6, and 7.
The electronic structure of the $n^{\rm th}$ alkali atom can be roughly 
approximated by a single valence electron 
in the $nS$ state of a Coulomb field created by the 
closed shells of the inner electrons and the nucleus, which 
together have total electric charge $+1$.  Thus the energy scale 
of the first electronic excitation of the Rb atom can be estimated 
from the hydrogen spectrum:  
$E_{\rm electronic} \approx E_{\rm Rydberg}/n^2$,
where $E_{\rm Rydberg} = 13.6$ eV.

An alkali atom in its electronic ground state has multiple spin states.
There are two contributions to its spin:
the electronic spin $\bm{S}$ with quantum number $s={1\over2}$
and the nuclear spin $\bm{I}$ with quantum number $i$.
The $2(2i+1)$ spin states can be labeled $|m_s,m_i \rangle$,
where $m_s$ and $m_i$ specify the eigenvalues of $S_z$ and $I_z$.
The Hamiltonian for a single atom includes a {\it hyperfine term}
that can be expressed in the form
\begin{eqnarray}
H_{\rm hyperfine} =
{2 E_{\rm hf} \over (2 i + 1)\hbar^2} \bm{I} \cdot \bm{S}.
\label{Hhyperfine}
\end{eqnarray}
This term splits the
ground state of the atom into two hyperfine multiplets
with energies differing by $E_{\rm hf}$.
The eigenstates can be labeled by the eigenvalues of the
hyperfine spin $\bm{F} = \bm{I}+\bm{S}$.
The associated  quantum numbers $f$ and $m_f$
specify the eigenvalues of $\bm{F}^2$ and $F_z$.
The eigenvalues of $H_{\rm hyperfine}$ are
\begin{eqnarray}
E_{f,m_f} =
\frac{f(f+1) - i(i+1) - \frac{3}{4}}{2 i + 1} E_{\rm hf} .
\end{eqnarray}
The hyperfine multiplet consists of
$2i+2$ states with $f = i + {1\over2}$
and $2i$ states with $f = i - {1\over2}$.

In the presence of a magnetic field $\bm{B} = B \bm{\hat z}$,
the Hamiltonian for a single atom has a {\it magnetic term}.
The magnetic moment \mbox{\boldmath $\mu$} of the atom is dominated by the
term proportional to the spin of the electron:
\mbox{\boldmath $\mu$} = $\mu \, \bm{S}/({1\over2}\hbar)$.
The magnetic term in the Hamiltonian can be expressed in the form
\begin{eqnarray}
H_{\rm magnetic} =
- {2 \mu \over \hbar} \bm{S} \cdot \bm{B} .
\label{Hmagnetic}
\end{eqnarray}
If $B \neq 0$, this term
splits the two hyperfine multiplets into $2(2i+1)$ hyperfine states.
In a weak magnetic field satisfying $\mu B \ll E_{\rm hf}$,
each hyperfine multiplet
is split into $2f+1$ equally-spaced Zeeman levels $| f, m_f \rangle$.
In a strong magnetic field satisfying $\mu B \gg E_{\rm hf}$,
the states are split into a set of $2i+1$ states with $m_s=-{1\over2}$
whose energies increase linearly with $B$ and a set of $2i+1$ states with
$m_s=+{1\over2}$ whose energies decrease linearly with $B$.
Each of those states is the continuation in $B$ of a specific
hyperfine state $| f, m_f \rangle$ at small $B$.
It is convenient to label the states by the hyperfine quantum
number $f$ and $m_f$ for general $B$, in spite of the fact 
that those states are not eigenstates of $\bm{F}^2$ 
if $B \neq 0$.
We denote the eigenstates of $H_{\rm hyperfine} + H_{\rm magnetic}$
by $| f, m_f ; B \rangle$ and their eigenvalues by $E_{f,m_f}(B)$.
The two eigenstates with the maximal value of $|m_f|$
are independent of $B$:
\begin{eqnarray}
\left| f=i+\mbox{$1\over2$}, m_f= \pm (i+\mbox{$1\over2$}) ; B \right\rangle
= \left| m_s = \pm \mbox{$1\over2$}, m_i = \pm i \right\rangle.
\label{hfmax}
\end{eqnarray}
Their eigenvalues are exactly linear in $B$:
\begin{eqnarray}
E_{f,m_f}(B) = \frac{i}{2i+1} E_{\rm hf} \mp \mu B.
\end{eqnarray}
If $B \neq 0$, each of the other eigenstates $| f, m_f ; B \rangle$
is a linear superposition of the two states 
$| f= i -{1 \over 2}, m_f \rangle$
and $| f= i + {1 \over 2}, m_f \rangle$.

\begin{figure}[tb]
\centerline{\includegraphics*[width=10cm,angle=270,clip=true]{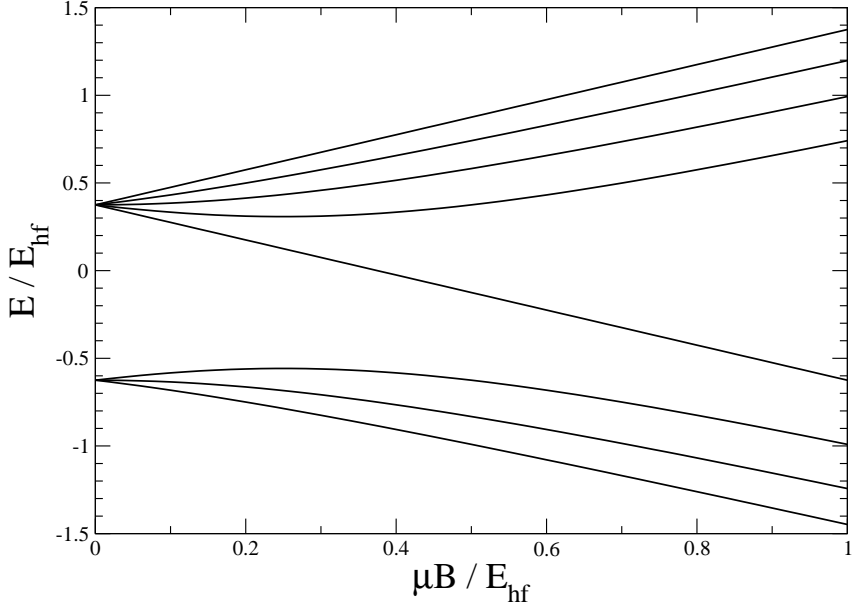}}
\caption{
The hyperfine energy levels as a function of the magnetic field 
for an alkali atom with $i = {3 \over 2}$, such as $^{87}$Rb.}
\label{fig:hyperfine}
\end{figure}

As an illustration, we take $^{87}$Rb atoms, whose nuclear spin 
quantum number is $i = {3 \over 2}$.  Since Rb is the $5^{\rm th}$
alkali atom, the energy scale of its first electronic excitation 
is approximately $E_{\rm Rydberg}/5^2 \approx 0.55$ eV.
At $B=0$, the electronic ground state is split into hyperfine 
multiplets with $f=1$ and $f=2$, with the $f=2$ multiplet higher in 
energy by $E_{\rm hf} = 2.83 \times 10^{-5}$ eV.
Thus the hyperfine splitting is more than 4 orders of magnitude 
smaller than the energy scale for electronic excitations.
The magnetic moment $\mu$ of a Rb atom
is approximately that of an electron:
$\mu \approx 2 \mu_B$, where $\mu_B$ is the Bohr magneton.
The magnetic energy scale $\mu B$ is comparable to 
$E_{\rm hf}$ when $B$ is about 2400 Gauss.
The dependence of the hyperfine energy levels on the magnetic field 
is illustrated in Fig.~\ref{fig:hyperfine}.

If the atoms have sufficiently low energy that
none of their electronic excitations can be excited,
they can be described as point particles with multiple spin states
that interact nonlocally through a potential.
In the case of two alkali atoms, their interactions at sufficiently
low energies can be described by the {\it Born-Oppenheimer potentials}
$V_s(r)$ and $V_t(r)$ for atoms whose valence electrons are
in {\it spin-singlet} and {\it spin-triplet} states, respectively.
The spin-dependent potential can be written as an outer-product operator
acting on the space of 2-atom spin states
$| m_s,m_i \rangle \otimes| m_s',m_i' \rangle$:
\begin{eqnarray}
{\mathcal V}(r) &=&
V_t(r) \left[ \mbox{$3\over4$} \, (1 \otimes 1)
    + \, (\mbox{$\sum_i$} S^i \otimes S^i)/\hbar^2 \right]
\nonumber
\\
&&+ V_s(r) \left[ \mbox{$1\over4$} \, (1 \otimes 1)
    - \, (\mbox{$\sum_i$} S^i \otimes S^i)/\hbar^2 \right].
\label{Vfun}
\end{eqnarray}
The asymptotic behaviors of the two potentials at large $r$ are
\begin{eqnarray}
V_t(r), \ V_s(r) & \longrightarrow & E_0 - {C_6 \over r^6},
\label{Vasymptotic}
\end{eqnarray}
where $E_0$ is the scattering threshold for two atoms
in the absence of the hyperfine and magnetic interactions.
The scattering thresholds are split by the hyperfine and
magnetic interactions in Eqs.~(\ref{Hhyperfine}) and (\ref{Hmagnetic}).
The threshold for two atoms in the states
$| f, m_f ; B \rangle$ and $| f', m_f' ; B \rangle$
is $2E_0 + E_{f,m_f}(B) +  E_{f',m_f'}(B)$.

There are also 3-body and higher $n$-body interactions between atoms.  
Their effects are generally believed to be negligible in most cases, 
and we will ignore them.  We will also ignore relativistic effects 
and retardation effects, which change the ultimate asymptotic 
behavior of the van der Waals potential from $1/r^6$ to $1/r^7$.

The nucleus of an atom with atomic number $Z$ and atomic mass number 
$A$ contains $Z$ protons and $A-Z$ neutrons.  The atom also contains 
$Z$ electrons to neutralize the electric charge.  Thus the total number 
of fermionic constituents in the atom is $A+Z$.
The atom is a boson if $A+Z$ is even and a fermion if 
$A+Z$ is odd.  In the case of alkali atoms, the atomic number $Z$ 
is always odd.  Thus an alkali atom is a boson if $A$ is odd 
and a fermion if $A$ is even.

If two atoms are in the same spin state $| f, m_f; B \rangle$,
their wavefunction $\psi(\bm{r}_1,\bm{r}_2)$ must satisfy 
a {\it symmetrization} condition.  If the atom is a boson, 
the wavefunction must be
symmetric under interchange of the two coordinates:
$\psi(\bm{r}_2,\bm{r}_1)= + \psi(\bm{r}_1,\bm{r}_2)$.
If the atom is a fermion, the wavefunction must be
antisymmetric under interchange of the two coordinates.
More generally, two atoms can be in superpositions of the 
various hyperfine states, in which case their 
wavefunction has multiple components 
$\psi_{m_{s1}m_{i1},m_{s2}m_{i2}}(\bm{r}_1,\bm{r}_2)$.
The symmetrization condition must be applied to the spin 
quantum numbers as well as to the coordinate dependence 
of the wavefunction.  If the atom is a boson, the components
of the wavefunction must satisfy
\begin{eqnarray}
\psi_{m_{s2}m_{i2},m_{s1}m_{i1}}(\bm{r}_2,\bm{r}_1) = 
+ \psi_{m_{s1}m_{i1},m_{s2}m_{i2}}(\bm{r}_1,\bm{r}_2).
\end{eqnarray}

We can now specify the fundamental theory.
The Hamiltonian is the sum of a one-body term for every particle
and a 2-body term for every pair of particles.
The one-body term is the sum of the kinetic energy $\bm{P}^2/(2m)$,
the hyperfine term in Eq.~(\ref{Hhyperfine}),
and the magnetic term in Eq.~(\ref{Hmagnetic}).
The 2-body potentials are given in Eq.~(\ref{Vfun}).
The bosonic or fermionic nature of the atoms is implemented
through constraints on the states.  If the atom is a boson, 
the $N$-particle wavefunction must be symmetric under interchange 
of any pair of atoms.
If the atom is a fermion, the $N$-particle wavefunction 
must be antisymmetric under such an interchange.

\subsection{Quantum field theory formulation}

The fundamental theory can be formulated as a 
{\it quantum field theory}.
This formulation involves $2(2i+1)$
quantum fields $\psi_{m_sm_i}(\bm{r})$  that can be arranged
into a column vector $\Psi(\bm{r})$.
The Hamiltonian is
\begin{eqnarray}
H_{\rm fun}  & = &
\int d^3r \, \left( {1 \over 2m} \nabla \Psi^\dagger \cdot \nabla \Psi
+ {2 E_{\rm hf} \over 2 i + 1}
    \Psi^\dagger \bm{I} \cdot \bm{S} \Psi
- 2 \mu  \bm{B} \cdot \Psi^\dagger \bm{S} \Psi \right)
\nonumber
\\
&& + {1 \over 2} \int d^3r_1 \int d^3 r_2 \,
    \left( \Psi(\bm{r}_1) \otimes \Psi(\bm{r}_2) \right)^\dagger
    {\mathcal V}(r_{12})
    \left( \Psi(\bm{r}_1) \otimes \Psi(\bm{r}_2) \right),
\end{eqnarray}
where $r_{12} = | \bm{r}_1 - \bm{r}_2 |$.
We have set $\hbar = 1$ to simplify the expression.
Dimensional analysis can be used to reintroduce $\hbar$
if desired.
The quantum fields $\psi_{m_s m_i} (\bm{r})$
satisfy equal-time commutation relations.
If the atom is a boson, the commutation relations are
\begin{subequations}
\begin{eqnarray}
\left[ \psi_{m_s m_i} (\bm{r}),
\psi_{m_s ' m_i '} (\bm{r}') \right] & = & 0,
\\
\left[ \psi_{m_s m_i} (\bm{r}),
\psi_{m_s ' m_i '} ^{\dagger} (\bm{r}') \right] & = &
\delta _{m_s m_s'}
\delta _{m_i m_i'}
\delta^3 (\bm{r} - \bm{r}').
\end{eqnarray}
\end{subequations}
If the atom is a fermion, the commutation relations are replaced
by anti-commutation relations.
The commutation relations imply that
the quantum field $\psi_{m_s m_i} (\bm{r})$
annihilates an atom in the spin state
$| m_i, m_s \rangle$ at the point $\bf r$.
They also enforce the constraint that quantum states
must be symmetric under interchange of two atoms in the same spin state.

The fundamental theory has several symmetries:
\begin{itemize}

\item {\it phase symmetry}.
The Hamiltonian $H_{\rm fun}$ is invariant under phase transformations:
$\Psi(\bm{r}) \to e^{i \theta}\Psi(\bm{r})$.
This symmetry is associated with the conservation of the number of atoms.

\item {\it translational symmetry}.
If the magnetic field $\bm{B}(t)$ is homogeneous,
$H_{\rm fun}$ is invariant under translations in space:
$\bm{r} \to \bm{r} + \bm{a}$.  This implies the conservation 
of the total momentum $\bm{P}$.

\item {\it time translational symmetry}.
If the magnetic field $\bm{B}(\bm{r})$ is static,
the theory is invariant under translations in time:
$t \to t + a$.  This implies the conservation 
of the total energy $E$, which is the eigenvalue of the 
Hamiltonian $H_{\rm fun}$.

\item {\it rotational symmetry}.
The Hamiltonian is invariant under rotations generated by the total 
angular momentum operator $\bm{J} = \bm{L} + \bm{S} + \bm{I}$,
where $\bm{L}$ is the orbital angular momentum, provided 
the magnetic field $\bm{B}$ is also transformed by the rotation.  
If $\bm{B} = 0$, this symmetry implies the conservation of the 
total angular momentum $\bm{J}$.
If $\bm{B} = B \bm{\hat z}$, it implies the conservation of the 
component $J_z$.

\item {\it parity symmetry}.
The Hamiltonian is invariant under the reflection
$\bm{r} \to - \bm{r}$.

\end{itemize}

The fundamental theory also has {\it Galilean symmetry}.
Since the Hamiltonian is the generator of translations in time,
the column vector of quantum fields can be extended to a 
time-dependent operator $\Psi(\bm{r},t)$:
\begin{eqnarray}
\Psi(\bm{r},t) =
e^{i H_{\rm fun} t} \Psi(\bm{r}) e^{-i H_{\rm fun} t}.
\end{eqnarray}
The transformation of this time-dependent operator under a 
Galilean boost with velocity vector $\bm{v}$ is
\begin{eqnarray}
\Psi(\bm{r},t) \longrightarrow
e^{i m \bm{v} \cdot \bm{r} - i ({1\over2} m v^2) t} 
\Psi(\bm{r} - \bm{v}t,t).
\label{Gal:fun}
\end{eqnarray}
This symmetry is not associated with a conservation law, 
because its generators do not commute with the Hamiltonian.  

If we restrict our attention to low energies with respect to a 
given scattering threshold, the fundamental theory has additional 
approximate symmetries.  
If the atoms have sufficiently low kinetic energy, their 
total orbital angular momentum $\bm{L}$ can be neglected.
If $\bm{B} = 0$, the total hyperfine spin $\bm{F} = \bm{I}+ \bm{S}$
is then conserved.
If there is a nonzero magnetic field $\bm{B} = B \bm{\hat z}$, 
the component $F_z$ is conserved.
The restriction on the energy may also impose a restriction 
on the possible spin states of the atoms.
In this case, the conservation of total atom number may be 
replaced by a more restrictive conservation law.  If $B=0$,
the energy restriction could allow only atoms in a specific 
hyperfine multiplet with quantum number $f$, in which case the 
total number of atoms in those $2f+1$ hyperfine states will be conserved.
If $B$ is large, the energy restriction may require all the atoms 
to be in the same spin state, in which case the number of atoms 
in that spin state will be conserved.

\subsection{Scattering}

If two atoms in the hyperfine states 
$| f_1, m_{f1}; B \rangle$ and $| f_2, m_{f2}; B \rangle$ collide,
they can either scatter elastically or they can
scatter into a different pair of hyperfine states
$| f_1', m_{f1}'; B \rangle$ and $| f_2', m_{f2}'; B \rangle$.
The scattering rate for each scattering channel
is determined by the $T$-matrix element,
which is a function of the relative wavenumbers
$\bm{k}$ and $\bm{k}'$ for the initial and final states.
If the collision energy relative to the scattering threshold
is sufficiently small,
elastic scattering is dominated by
the $S$-wave orbital angular momentum channel
and the $T$-matrix element
${\mathcal T}_{f_1 m_{f1},f_2 m_{f2}} (k)$
for elastic scattering reduces to a function of 
$k = |\bm{k}| = |\bm{k}'|$ only.
The low-energy limit of the $T$-matrix element is given by 
the scattering lengths for the hyperfine channel.
If the two hyperfine states are distinct, we denote the scattering 
length by $a_{f_1 m_{f1}, f_2 m_{f2}}$.  The low-energy limit 
of the $T$-matrix element is
\begin{eqnarray}
{\mathcal T}_{f_1 m_{f1}, f_2 m_{f2}} (k) \longrightarrow
- \frac{4 \pi}{m} a_{f_1 m_{f1}, f_2 m_{f2}}
\qquad  {\rm as} \: k \to 0.
\label{Tfmfm}
\end{eqnarray}
If the two atoms are in the same hyperfine state $| f, m_f; B \rangle$,
we denote the scattering length by $a_{f m_{f}}$.  The low-energy limit 
of the $T$-matrix element is
\begin{eqnarray}
{\mathcal T}_{f m_{f}, f m_{f}} (k) \longrightarrow
- \frac{8 \pi}{m} a_{f m_{f}}
\qquad  {\rm as} \: k \to 0.
\label{Tfm}
\end{eqnarray}
If the atoms are fermions, the scattering length $a_{f m_{f}}$ 
vanishes because the wavefunction for an S-wave state must be 
symmetric under interchange of the coordinates. 
The cross section for the scattering of distinct hyperfine states 
are obtained by squaring the T-matrix element
${\mathcal T}_{f_1 m_{f1}, f_2 m_{f2}} (k)$ and 
multiplying it by the flux factor $m/(2 k)$ 
and by the phase space factor $m k/(2\pi)$.  
If the two atoms are in the same hyperfine state, 
the phase space factor is multiplied by a factor of ${1 \over 2}$
to compensate for the double-counting of the states 
of the identical particles.  Thus the low-energy limit of the 
cross section for the elastic scattering of two atoms in the 
$|f,m_f;B \rangle$ hyperfine state is
\begin{eqnarray}
\sigma_{\rm elastic}(E) \longrightarrow
8 \pi |a_{f m_{f}}|^2
\qquad  {\rm as} \: E\to 0.
\label{sigmafm}
\end{eqnarray}

\begin{figure}[tb]
\centerline{\includegraphics*[width=10cm,angle=270,clip=true]{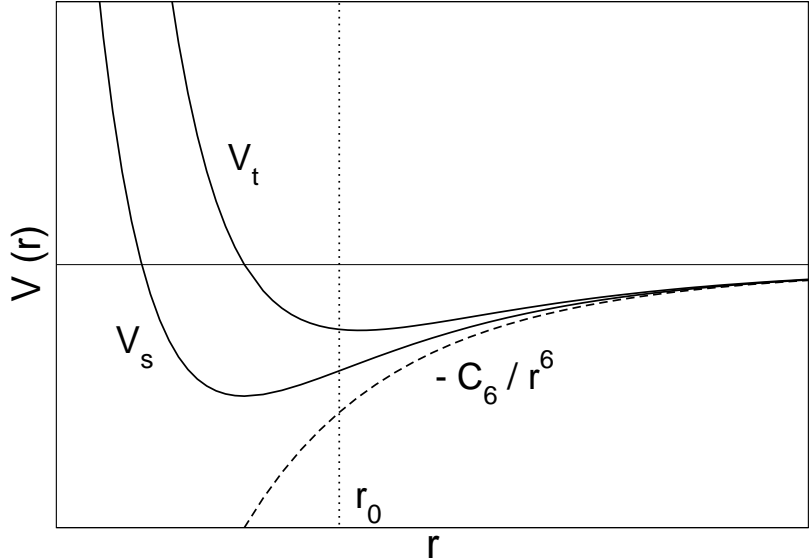}}
\caption{
Qualitative behavior of the spin-singlet potential 
(lower solid line) and the spin-triplet potential 
(upper solid line).  The low-energy scattering properties 
can be reproduced by replacing both potentials by the
van der Waals potential $-C_6/r^6$ with a boundary condition 
at $r=r_0$ such that the spin-singlet and spin-triplet 
scattering lengths have the correct values $a_s$ and $a_t$.}
\label{fig:spinpots}
\end{figure}

In the absence of the hyperfine and magnetic interactions, 
scattering through the spin-singlet and spin-triplet potentials 
in Eq.~(\ref{Vfun}) determines the spin-singlet 
and spin-triplet scattering lengths $a_s$ and $a_t$.
The spin-triplet scattering length $a_t$
is the scattering length for a pair of atoms that are both in the
same hyperfine state given in Eq.~(\ref{hfmax}).
The only effect of the hyperfine and magnetic terms 
on this channel is to change the scattering threshold.
If it were not for the hyperfine term in Eq.~(\ref{Hhyperfine}),
the scattering lengths for all the other hyperfine channels
could be expressed as linear combinations of
$a_s$ and $a_t$ using Clebsch-Gordon coefficients.
Because of the hyperfine term, it is necessary to solve the
coupled-channel Schr\"odinger equation numerically to determine
the scattering lengths for all the other hyperfine channels.
However, if the scattering lengths $a_s$ and $a_t$ are known,
there is a simple approximation to the coupled-channel problem
that reproduces the low-energy scattering observables,
including the scattering lengths, quite accurately \cite{JG06}.
As illustrated in Fig.~\ref{fig:spinpots},
the spin-singlet and spin-triplet potentials
$V_s(r)$ and $V_t(r)$ in Eq.~(\ref{Vfun}) can be replaced by their
asymptotic forms in Eq.~(\ref{Vasymptotic}) together with boundary
conditions at a small separation $r_0$. The boundary
conditions are tuned so that the spin-singlet and spin-triplet
scattering lengths have the desired values $a_s$ and $a_t$
for $B=0$ and $E_{\rm hf}=0$.
Those same boundary conditions are then used for nonzero values 
of $B$ and $E_{\rm hf}$.
Thus the low-energy scattering problem for all the hyperfine 
channels is essentially determined by just 5 parameters:
$a_s$, $a_t$, $C_6$, $E_{\rm hf}$, and $\mu B$.

We now focus on a specific hyperfine channel.
To be definite, we take the alkali atom to be a boson.
We also take the two atoms to be in the same hyperfine state
$| f, m_f ; B \rangle$,
so they are identical bosons.  We choose the origin of energy
to be the  scattering threshold for that channel:
$2E_0 + 2 E_{f,m_f}(B)$.
The T-matrix element for S-wave scattering of two particles
with collision energy $E = k^2/m$ can be expressed in the form
\begin{eqnarray}
{\mathcal T}(k) =  
{8 \pi \over m} {1 \over k \cot \delta_0(k) - i k} ,
\label{T-delta}
\end{eqnarray}
where $\delta_0(k)$ is the S-wave phase shift.
If there are no lower hyperfine channels into which the atoms can
scatter, the unitarity of the S-matrix implies that the
phase shift is real valued.
The low-energy behavior of the phase shift
is given by the {\it effective range expansion}:
\begin{eqnarray}
k \cot \delta_0(k) = 
- 1/a + \mbox{$1\over 2$} r_s k^2 + \ldots ,
\label{effrangeexp}
\end{eqnarray}
which defines the {\it scattering length} $a=a_{f m_f}$
and the {\it effective range} $r_s$.
Taking the limit $k \to 0$ in Eq.~(\ref{T-delta})
and using Eq.~(\ref{effrangeexp}), 
we recover the limiting expression for the T-matrix element
in Eq.~(\ref{Tfm}).

If the fundamental theory had only short-range potentials, 
$k \cot\delta_0(k)$ would be an analytic function of $k^2$
at $k=0$ and it would therefore have a power series 
expansion in $k^2$.  However, the asymptotic behavior 
of the potentials $V_s(r)$ and $V_t(r)$,
as given in Eq.~(\ref{Vasymptotic}), is $1/r^6$.
As a consequence, beginning at order $k^4$, there are 
logarithms of $k$ in the coefficients of the powers of $k$.
Thus $k \cot\delta_0(k)$ has a well-behaved expansion 
in powers of $k^2$ only to order $k^2$.  
Thus the scattering length and
the effective range defined by Eq.~(\ref{effrangeexp})
are the only well-defined coefficients 
in the effective range expansion for atoms.

The important energy scales for low-energy alkali atoms 
include the hyperfine splitting $E_{\rm hyperfine}$, 
the magnetic energy scale $\mu B$,
and the van der Waals energy scale defined by
\begin{eqnarray}
E_{\rm vdw} = (m^3 C_6 / \hbar ^6) ^{-1/2} .
\label{EvdW}
\end{eqnarray}
For Rb atoms, the van der Waals energy scale is 
$E_{\rm vdW} = 6.4 \times 10^{-9}$ eV.
This is more than 3 orders of magnitude smaller than 
the hyperfine splitting $E_{\rm hf}$.
The magnetic energy scale $\mu B$ is comparable to 
$E_{\rm vdW}$ when the magnetic field is about 0.54 Gauss.
The van der Waals energy scale corresponds to a temperature of 
$7.4 \times 10^{-5}$ K.  We consider atoms to be {\it ultracold} 
if their kinetic energies are much smaller
than the van der Waals energy scale.
In subsequent sections, we will review scattering models 
that can describe such ultracold atoms.

\subsection{Large scattering length}
\label{sec:large}

The natural scale for the scattering length $a$ of a specific 
hyperfine state is the van der Waals length defined by
\begin{eqnarray}
\ell_{\rm vdw} = (m C_6 / \hbar ^2) ^{1/4} .
\label{ellvdW}
\end{eqnarray}
A scattering length can be orders of magnitude larger than 
$\ell_{\rm vdw}$ only if there is a fine-tuning of some parameter 
in the Hamiltonian.
If $|a|$ is much larger than $\ell_{\rm vdw}$, 
we refer to $a$ as a {\it large scattering length}.
If $a$ is large, ultracold atoms have an important energy scale
that is much smaller than the van der Waals energy scale
$E_{\rm vdw}$, namely $\hbar^2/(m a^2)$.

For particles with short-range interactions, the scattering length $a$
is considered to be large if its absolute value is much larger than  
the range.  Particles with large scattering lengths have universal 
low-energy properties that depend on the scattering length but are 
insensitive to other details of the interactions.
The known universal properties have been summarized in a recent 
review \cite{Braaten:2004rn}.  In the case of atoms, the relevant 
range is the van der Waals length scale $\ell_{\rm vdw}$
defined in Eq.~(\ref{ellvdW}).  In the 2-atom sector,
the universal properties are rather simple.
The cross section for elastic scattering of bosonic atoms in the 
same spin state with energy in the region $|E| \ll E_{\rm vdW}$,
where $E_{\rm vdW}$ is the van der Waals energy scale defined 
in Eq.~(\ref{EvdW}), is given by the universal formula
\begin{eqnarray}
\sigma_{\rm elastic}(E) = \frac{8 \pi a^2}{1 + m a^2 E}.
\label{sigel-uni}
\end{eqnarray}
Shallow bound states whose binding energies are much smaller than 
$E_{\rm vdW}$ are also universal.
If $a<0$, there are no such bound states.
If $a>0$, there is a single shallow bound state
with binding energy
\begin{eqnarray}
E_D = 1/(m a^2) .
\label{E-dimer}
\end{eqnarray}
We will refer to this state as the {\it shallow dimer}.

The natural scale for the scattering lengths $a_s$ and $a_t$
is the van der Waals length defined in Eq.~(\ref{ellvdW}).
The scattering length $a_s$ or $a_t$ can be much larger than 
$\ell_{\rm vdw}$ only if the 
depth of the potential $V_s(r)$ or $V_t(r)$ is tuned so that 
there is a bound state very close to the threshold.
If $a_s$ or $a_t$ is large, several of the scattering lengths 
$a_{f m_f}$ of the hyperfine channels will also be large.
Examples of alkali atoms that have large 
scattering lengths at zero magnetic field are 
$^6$Li ($a_t \approx -2160~a_0$), 
$^{85}$Rb ($a_s \approx + 2800~a_0$), 
and $^{133}$Cs ($a_t \approx +2400~a_0$).

A large scattering length $a_{f m_f}$ in a specific hyperfine 
channel can be obtained by tuning the magnetic field $B$ to a 
{\it Feshbach resonance} $B_i$.  
If the magnetic field is near $B_i$, there is a diatomic molecule 
near the scattering threshold for two atoms in the 
$| f, m_f; B \rangle$ hyperfine state.
The diatomic molecule is a bound state of atoms in a spin state 
with a higher threshold.  Since the magnetic moment of the 
diatomic molecule differs from twice the magnetic moment of the 
$| f, m_f; B \rangle$ atom,  
the energy of the molecule can be made to cross the scattering 
threshold by changing the magnetic field $B$.
The {\it Feshbach resonance} occurs at the value $B_i$ for 
which the molecule is at resonance with two atoms at the 
scattering threshold.
The existence of Feshbach  resonances in Cs atoms was first pointed out 
by Tiesinga, Verhaar, and Stoof \cite{TVS93}. 
The first Feshbach resonances to be observed were the
resonances at 853~G and 907~G for $^{23}$Na atoms in the $|1, -1 \rangle$
hyperfine level \cite{Inouye98}
and the resonance at 155~G for $^{85}$Rb atoms in the
$|2, -2 \rangle$ hyperfine level \cite{Courteille98}.
The physics of Feshbach resonances in cold atoms 
has been summarized in a recent review \cite{KGP06}.

At a generic value $B_0$ of the magnetic field,
the scattering length $a(B)$ for a specific hyperfine channel
varies slowly as a function of the magnetic field $B$.  It can 
therefore be expanded as a power series around $B=B_0$:
\begin{eqnarray}
a(B) = a(B_0) + a'(B_0) (B-B_0) + \dots .
\label{a-smooth}
\end{eqnarray}
At a Feshbach resonance, the scattering length is not a smooth function 
of $B$.  When the magnetic field
is well below the Feshbach resonance $B_i$, 
the scattering length has some off-resonant value
$a_{\rm bg}$. As $B$ increases through the value $B_i$, 
$a(B)$ diverges to $+ \infty$ or $-\infty$, jumps discontinuously to 
$- \infty$ or $+\infty$, and then approaches the original
off-resonant value $a_{\rm bg}$.  
The scattering length must pass through
zero at some value of $B$, either before or after $B_i$.
If we denote that value by $B_i + \Delta_i$, the
scattering length can be expressed as
\begin{eqnarray}
a(B) = a_{\rm bg}(B) \left( 1 - \frac {\Delta_i} {B-B_i} \right) ,
\label{a:FR}
\end{eqnarray}
where $a_{\rm bg}(B)$ is a smooth function of the magnetic field 
that can be expanded as a power series around $B_i$ as in 
Eq.~(\ref{a-smooth}). The
parameters $B_i$ and $\Delta_i$ are convenient, because it is often
possible to measure the locations of divergences
and zeroes of the scattering length
more accurately than its value $a(B)$ at any particular value of $B$.


\section{Scattering models}
\label{sec:scatmod}

The fundamental theory described in Sec.~\ref{sec:fun}
provides an extremely accurate description of systems
consisting of atoms in any of the hyperfine states
as long as their energies are small compared to the electronic
excitation energy of the atom.
If we restrict our attention to ultracold atoms,
whose energies relative to a specific scattering threshold
are small compared to the van der Waals energy 
scale $E_{\rm vdW}$ defined in Eq.~(\ref{EvdW}),
we can describe the system accurately
using a much simpler model.  The effects of virtual
atoms whose energies are far from the 
scattering threshold can be taken into account through the 
parameters of the model.

We take the atoms of primary interest to be those
in a specific hyperfine state 
$|f,m_f;B \rangle$. We choose the zero of energy to be the 
scattering threshold for atoms in that hyperfine state.
We restrict our attention to ultracold atoms whose energies 
satisfy $|E| \ll E_{\rm vdW}$.
If $\mu B \ll E_{\rm vdw}$, this energy constraint can 
be satisfied by atoms in any of the $2f+1$ spin states 
in the same hyperfine multiplet.
For larger values of $B$, only atoms in the single 
hyperfine state $|f,m_f;B \rangle$ satisfy this energy constraint.

\subsection{Potential models}

One class of models that can be used to describe the  
behavior of ultracold atoms in the spin state of interest is
{\it potential models}.
A potential model is specified by interaction potentials 
between the constituents of the model.
In the simplest potential models, the only constituents are atoms 
in the spin state of interest.  
In this case, the potential model 
is specified by a single potential $V(r)$.
The potential $V(r)$ must be tuned so that the solutions to the 
single-channel Schr\"odinger equation for that potential 
give the same results for low-energy scattering cross 
sections and low-energy bound states
as the solution to the coupled-channel Schr\"odinger equation 
for the fundamental theory.
The potential $V(r)$ need not resemble either of the potentials
$V_s (r)$ or $V_t (r)$ in the fundamental theory as long as 
it accurately reproduces low-energy observables for 
atoms in the spin state of interest.
If we restrict our attention to sufficiently low energies
that the scattering length $a$ is the only relevant interaction 
parameter, then all that is required is that the potential
be short-ranged and give the correct scattering length.
 
A potential model can be formulated as a quantum field theory.
If the only constituents in the model are atoms in a single 
hyperfine state $| f, m_f;B \rangle$, there is a single 
quantum field $\psi(\bm{r})$ which
annihilates atoms in that hyperfine state.
The Hamiltonian is
\begin{eqnarray}
H_{\rm pot}  & = &
\int d^3r \, {1 \over 2m} \nabla \psi^\dagger \cdot \nabla \psi
\nonumber
\\
&& + {1 \over 2} \int d^3r_1 \int d^3 r_2 \,
    \psi(\bm{r}_1)^\dagger \psi(\bm{r}_2)^\dagger 
    V(r_{12}) \psi(\bm{r}_1) \psi(\bm{r}_2) .
\end{eqnarray}
The time-dependent quantum field $\psi(\bm{r},t)$
satisfies equal-time commutation relations.
If the atom is a boson, the commutation relations are
\begin{subequations}
\begin{eqnarray}
\left[ \psi(\bm{r},t), \psi(\bm{r}',t) \right] & = & 0,
\\
\left[ \psi(\bm{r},t), \psi^{\dagger}(\bm{r}',t) \right] & = &
\delta^3 (\bm{r} - \bm{r}').
\end{eqnarray}
\end{subequations}

A more elaborate potential model could describe $N-1$ additional
spin states of the atoms.  The model would be specified 
by the diagonal potentials for each pair of spin states 
and by coupling potentials between pairs of spin states.
These potentials need not resemble those of the 
fundamental theory
as long as the solutions to the Schr\"odinger equation 
for all the coupled channels reproduce
the predictions of the fundamental theory for low-energy 
observables involving the spin state of interest.

\subsection{Scattering Models}

Another class of models that can be used to describe the low-energy 
behavior of the atoms is {\it scattering models}.
A scattering model can be specified by the
$T$-matrix elements for scattering of the constituents 
of the model along with any additional parameters 
that are required to make the model well-defined.
In the simplest scattering models, the only constituents are atoms 
in a single spin state.  At sufficiently low energies, 
the atoms have only S-wave scattering and the T-matrix element 
reduces to a function ${\mathcal T}(k)$ of the magnitude of the
relative momentum $k$ only. 
Since it is specified in terms of the low-energy 
observables ${\mathcal T}(k)$, a scattering model provides a particularly 
natural description for ultracold atoms.

A more elaborate scattering model could include other spin states
of the atom as additional degrees of freedom. 
In this case, the T-matrix elements for all the pairs of spin states
could be part of the specification of the model.
In contrast to potential models, diatomic molecules can also be
included as explicit degrees of freedom in a scattering model.
In this case, T-matrix elements involving the diatomic 
molecules could also be part of the specification of the model.

\subsection{Hamiltonian quantum field theories}

A scattering model can be formulated as a quantum 
field theory. If the T-matrix element ${\mathcal T}(k)$ 
that is part of the specification of the scattering model
is an analytic function of $k^2$ at $k=0$, the T-matrix element
can be reproduced by a {\it local quantum field theory}.
A local quantum field theory is one whose interactions only 
involve the quantum fields and their derivatives at the same point 
in space.  If a local quantum field theory has a Hamiltonian 
formulation, its Hamiltonian can be expressed as the integral 
over space of a Hamiltonian density: 
\begin{eqnarray}
H = \int d^3 r \left( {\mathcal H}_{\rm free} + {\mathcal H}_{\rm int} \right).
\label {Hlocal}
\end{eqnarray}
If the quantum field $\psi(\bm{r})$ that annihilates atoms 
in the spin state of interest is the only quantum field
in the model, the free term in the Hamiltonian density is
\begin{eqnarray}
{\mathcal H}_{\rm free} & = &
{1 \over 2m} \nabla \psi^\dagger \cdot \nabla \psi .
\label {H:free}
\end{eqnarray}
The interaction term in the Hamiltonian density must be a polynomial
in $\psi$ and $\psi^\dagger$ and their gradients, such as
\begin{eqnarray}
{\mathcal H}_{\rm int} & = &
{1 \over 4} \lambda_0\left( \psi^\dagger \right)^2 \psi^2
\nonumber
\\
&& + {1\over 4} \rho_0 
\left( (\psi^\dagger)^2 \nabla \psi \cdot \nabla \psi
+ 2 \psi^\dagger \psi \nabla \psi^\dagger \cdot \nabla \psi
+ \psi^2 \nabla \psi^\dagger \cdot \nabla \psi^\dagger \right)
+ \ldots .
\label {H:int}
\end{eqnarray}

One advantage of a Hamiltonian quantum field theory is that it 
guarantees that the space of quantum states is a Hilbert space.
Since the Hamiltonian $H$ is a hermitian operator, 
its eigenstates $| n \rangle$ form a complete set of states. 
The completeness relation can be expressed formally as
$\sum_n | n \rangle \langle n | = 1$.  If we express the 
S-matrix in the form ${\mathcal S}= 1 + i {\mathcal T}$, 
the completeness relation can be used to express the
unitarity condition ${\mathcal S}^\dagger {\mathcal S} = 1$
in the form
\begin{eqnarray}
2 \, {\rm Im} \, \langle i | {\mathcal T} | i \rangle =
\sum_n | \langle n | {\mathcal T} | i \rangle |^2 .
\label{optical}
\end{eqnarray}
This form of the unitarity condition is called the 
{\it Optical Theorem}.   Note that the right side is the sum 
of positive-definite terms.

\subsection{Lagrangian quantum field theories}

It is sometimes useful to consider a more general class 
of quantum field theories that have a Lagrangian formulation 
but do not necessarily have a
Hamiltonian formulation.  The Lagrangian for a local quantum field 
theory can be expressed as the integral over space of a Lagrangian
density: 
\begin{eqnarray}
L = 
\int d^3 r \left( {\mathcal L}_{\rm free} + {\mathcal L}_{\rm int} \right).
\label {Llocal}
\end{eqnarray}
If the quantum field $\psi(\bm{r})$ that annihilates atoms 
in the spin state of interest is the only quantum field
in the model, the free term in the Lagrangian density is
\begin{eqnarray}
{\mathcal L}_{\rm free} =
\frac{1}{2} i \left(
\psi ^\dagger \frac{\partial}{\partial t} \psi
- \frac{\partial}{\partial t} \psi ^\dagger \psi
\right)
- \frac{1}{2m}
\nabla \psi ^\dagger \cdot \nabla \psi .
\label{Lfree:real}
\end{eqnarray}
This form is manifestly real. If we ignore terms
that are total time derivatives or total divergences,
it can be written in a more compact form:
\begin{eqnarray}
{\mathcal L}_{\rm free} = \psi ^\dagger \left(
i \frac{\partial}{\partial t}
+ \frac{\nabla ^2} {2 m}
\right) \psi  .
\label{Lfree}
\end{eqnarray}
If the quantum field theory has a Hamiltonian 
formulation, the interaction term in the Lagrangian density is simply
${\mathcal L}_{\rm int} = - {\mathcal H}_{\rm int}$.
However, Lagrangian quantum field theories are a more general class 
of models than Hamiltonian quantum field theories.
They can, for example, have interaction terms that involve
time derivatives, such as
\begin{eqnarray}
{\mathcal L}_{\rm int} = \frac{1}{4} m \lambda_0 '
(\psi^2)^\dagger   
\left( i \frac{\partial} {\partial t} + \frac{\nabla^2}{4m} \right) 
\psi^2 + \ldots.
\label{Lint:timederiv}
\end{eqnarray}
More generally, ${\mathcal L}_{\rm int}$ must be a polynomial in $\psi$,
$\psi^\dagger$, their gradients, and their time-derivatives.

If all we have is a Lagrangian formulation of the quantum field theory, 
there is no guarantee that the Optical Theorem will have the
form in Eq.~(\ref{optical}), where 
${\rm Im} \, \langle i | {\mathcal T} | i \rangle$
is expressed as the sum of positive-definite terms.  
The general expression may have the form
\begin{eqnarray}
2 \, {\rm Im} \, \langle i | {\mathcal T} | i \rangle =
\sum_n | \langle n | {\mathcal T} | i \rangle |^2 
- \sum_m | \langle m | {\mathcal T} | i \rangle |^2 ,
\label{optical:+-}
\end{eqnarray}
where the states labeled $| n \rangle$ and $| m \rangle$
together form a complete set.  One way to interpret 
such a relation is that the space of quantum states 
is a complex vector space with indefinite norm.
The completeness relation for this vector space has the form
$\sum_n |n \rangle \langle n| - \sum_m |m \rangle \langle m|= 1$. 
The states $| m \rangle$ can be interpreted as states 
with {\it negative norm}.

The problem of negative-norm states
arises in covariant formulations of Quantum Electrodynamics.
However, one can use gauge invariance to show that there is a 
{\it physical subspace} of the complex vector space spanned 
by the positive-norm states $| n \rangle$ such that if 
$| i \rangle$ is in the physical subspace, 
${\rm Im} \, \langle i | {\mathcal T} | i \rangle$
can be written as the sum of positive-definite terms:
\begin{eqnarray}
2 \, {\rm Im} \, \langle i | {\mathcal T} | i \rangle =
\sum_{{\rm physical} \, n} | \langle n | {\mathcal T} | i \rangle |^2.
\label{optical:phys}
\end{eqnarray}
This requires a cancelation between positive and negative terms 
in Eq.~(\ref{optical:+-}). 
In QED, this cancelation is guaranteed by gauge invariance.
In a scattering model for atoms, there is no obvious mechanism 
to guarantee such a cancelation.  Thus the existence of 
negative-definite terms in the imaginary part of a diagonal 
T-matrix element is more problematic.

\subsection{Symmetries}
 
The symmetries of the fundamental theory can be used to constrain the
Hamiltonian or the Lagrangian of a local quantum field theory.
We will give the constraints only for the simple case of a single 
quantum field $\psi(\bm{r},t)$:
\begin{itemize}

\item {\it phase symmetry}.
The phase transformation is
$\psi(\bm{r},t) \to e^{i \theta}\psi(\bm{r},t)$.
Invariance under this transformation requires that every term in 
${\mathcal H}_{\rm int}$ or ${\mathcal L}_{\rm int}$
have an equal number of factors of $\psi$ and $\psi^\dagger$.

\item {\it translational symmetry}.
This is implemented automatically in a local quantum field theory 
for which ${\mathcal H}_{\rm int}$ or ${\mathcal L}_{\rm int}$ is a function of
$\psi$, $\psi^\dagger$, their gradients, and their time-derivatives.

\item {\it time translational symmetry}.
This is also implemented automatically in a local quantum field theory.

\item {\it rotational symmetry}.
If the model involves only a single quantum field $\psi(\bm{r},t)$,
a rotation acts only on its vector argument $\bm{r}$.
The requirement of rotational symmetry is that
${\mathcal H}_{\rm int}$ or ${\mathcal L}_{\rm int}$  be a scalar under rotations.  
This requires each gradient $\nabla^i$ to have its index $i$ contracted 
with that of another gradient $\nabla^i$
or with a Levi-Civita tensor $\epsilon^{ijk}$.

\item {\it parity symmetry}. 
This can be used to exclude terms in ${\mathcal H}_{\rm int}$ 
or ${\mathcal L}_{\rm int}$ with a Levi-Civita tensor.

\end{itemize}

{\it Galilean symmetry} imposes strong constraints 
on the interaction terms 
in ${\mathcal H}_{\rm int}$ or ${\mathcal L}_{\rm int}$. 
The Galilean transformation with velocity vector $\bm{v}$
in the fundamental theory is given in Eq.~(\ref{Gal:fun}).
The corresponding Galilean transformation for the field 
$\psi(\bm{r},t)$ is
\begin{eqnarray}
\psi(\bm{r},t) \longrightarrow
e^{i m \bm{v} \cdot \bm{r} - i ({1\over2} m v^2) t} 
\psi(\bm{r} - \bm{v}t,t).
\label{Gal:mod}
\end{eqnarray}

One way to construct interaction terms that are Galilean invariant 
is to construct them out of building blocks with simple behavior 
under Galilean transformations.  One simple building block  
is the product $\psi^\dagger \psi$, which is Galilean invariant.  
Thus terms constructed out of 
$\psi^\dagger \psi$ and its gradients, such as
$\nabla^i(\psi^\dagger \psi)$ and 
$\nabla^i \nabla^j(\psi^\dagger \psi)$, 
are Galilean invariant.
The first term in ${\mathcal H}_{\rm int}$ in Eq.~(\ref{H:int})
is Galilean invariant, because it can be expressed as  
$(\psi^\dagger \psi)^2$.  The second term 
in ${\mathcal H}_{\rm int}$ is also Galilean invariant,
because it can be expressed in the form
$\nabla (\psi^\dagger \psi) \cdot \nabla (\psi^\dagger \psi)$.

Another simple building block is 
$\psi {\stackrel \leftrightarrow \nabla}{}^2 \psi$, 
where the operator
$\stackrel \leftrightarrow \nabla{}^2$ is defined by
\begin{eqnarray}
\chi {\stackrel \leftrightarrow \nabla}{}^2 \psi =
\chi  (\nabla^2 \psi) - 2 (\nabla \chi) \cdot (\nabla \psi) 
+ (\nabla^2 \chi) \psi.
\label{nabla^2}
\end{eqnarray}
The operator $\psi {\stackrel \leftrightarrow \nabla}{}^2 \psi$ 
transforms under Galilean 
transformations in the same way as $\psi^2$.
A further set of simple building blocks is 
the combination 
$[i \frac{\partial} {\partial t} + \nabla^2/(2Nm)]\psi^N$,
where $N = 1,2,\ldots$.  They transform under Galilean 
transformations in the same way as $\psi^N$.
This combination with $N=1$ appears in the expression for 
${\mathcal L}_{\rm free}$ in Eq.~(\ref{Lfree}).
This combination with $N=2$ appears in the interaction term 
in Eq.~(\ref{Lint:timederiv}).

If the energy restriction allows atoms 
in multiple spin states related by a symmetry,
the symmetry can be imposed on the corresponding quantum fields 
$\psi_m(\bm{r},t)$.  For example, if $B = 0$, 
the $2f+1$ spin states in a hyperfine multiplet 
form an irreducible $(2f+1)$-dimensional representation 
of the $SU(2)$ symmetry group of angular momentum.   
There is also a $U(1)$ symmetry corresponding to
multiplying all the fields $\psi_m(\bm{r},t)$ by the same phase.

\subsection{Two-body observables}

In a quantum field theory, physical observables can be conveniently
encoded in the correlation functions of the theory.
The physical observables in the 2-atom sector can be encoded in the
amputated connected Green's function ${\mathcal A}$ for two atoms 
in the asymptotic past
to evolve into two atoms in the asymptotic future.  In general, this
amplitude is a function of the energies $E_1$ and $E_2$ and the momenta
$\bm{p}_1$ and $\bm{p}_2$ of the two incoming atoms and 
the energies $E_1'$ and $E_2'$ and the momenta $\bm{p}_1'$ and 
$\bm{p}_2'$ of the two outgoing atoms. 
The atoms can be off their energy shells: $E_i$ need not be equal to
$p_i^2/(2m)$.
The total energy $E$ and the total momentum $\bm{P}$ are conserved:
$E_1 + E_2 = E_1' + E_2'=E$ and
$\bm{p}_1 + \bm{p}_2 = \bm{p}_1' + \bm{p}_2' = \bm{P}$.
Galilean invariance constrains the amplitude to be a function of
$E-\bm{P}^2/(4m)$ and the relative momenta
$\bm{k} = {1\over 2}(\bm{p}_1 - \bm{p}_2)$
and $\bm{k}' = {1\over 2}(\bm{p}_1' - \bm{p}_2')$.  We can simplify the
problem by working in the center-of-mass frame $\bm{P}=0$.
If the energy is restricted to be low enough that
there is S-wave scattering only, the amplitude in the center-of-mass frame
depends only on the magnitudes of $\bm{k}$ and $\bm{k}'$,
so it reduces to a function ${\mathcal A}(E;k,k')$ of three variables.

The amplitude ${\mathcal A}(E;k,k')$ encodes all information 
about low-energy 2-atom scattering.
The T-matrix element ${\mathcal T}(k)$ for the S-wave scattering of
two atoms with relative momentum $\bm{k}$ into two atoms with
relative momentum $\bm{k}'$
is obtained by evaluating ${\mathcal A}(E;k,k')$ at the on-shell point
obtained by setting all the atoms on their energy shells:
$E  =  2 (k^2/2 m) = 2  ({k'}^2/2 m)$,
which requires $k = k'$.  The T-matrix element is
\begin{eqnarray}
{\mathcal T}(k)   =  {\mathcal A}(k^2/m;k,k) .
\end{eqnarray}
The cross section for the elastic scattering of two 
identical bosonic atoms with total kinetic energy
$E= k^2/m$ is obtained by squaring the T-matrix element and
multiplying by the phase space factor $m k/(4\pi)$ 
and by the flux factor $m/(2 k)$:
\begin{eqnarray}
\sigma(E)  =  {m^2 \over 8 \pi} \left| {\mathcal T}(k) \right|^2 .
\end{eqnarray}

Information about S-wave bound states is also encoded in
${\mathcal A}(E;k,k')$ through its poles in the energy $E$.
The amplitude ${\mathcal A}(E;k,k')$ is a double-valued function of
the energy variable $E$, which takes its values on the
two-sheeted complex plane with a branch point at $E=0$
and a branch cut along the positive $E$ axis.
The interpretation of the pole in $E$ depends on
where the pole resides in the complex $E$ plane.
For simplicity, we assume that the atom has no spin states 
with lower energy.
A pole on the negative real axis of the physical sheet
of the complex energy $E$ corresponds to a {\it stable bound state}.
A pole on the negative real axis of the unphysical sheet 
is called a {\it virtual state}.
A pole on the unphysical sheet just below the positive real axis
corresponds to a {\it scattering resonance}.

\subsection{Renormalization}
\label{sec:renorm}
 
One complication of local quantum field theories is that they are 
inherently singular at short distances.
The theory is well-defined only if it is modified in some way 
to suppress the effects of particles that approach
arbitrarily close to each other or that have arbitrarily high energy.
A convenient way to do this is to impose an upper limit on the 
wavevector of virtual particles: $|\bm{k}| < \Lambda$.
The variable $\Lambda$ is called the {\it ultraviolet cutoff}.
The singularities then appear in the limit $\Lambda \to \infty$.
However, it is possible to avoid the singularities by using 
a procedure called {\it renormalization}.  The parameters 
of the model, such as $\lambda_0$ and $\rho_0$ in Eq.~(\ref{H:int}), 
are made functions of $\Lambda$ 
in such a way that the model has a well-behaved and nontrivial 
limit as $\Lambda \to \infty$.

There are two rather different approaches to renormalization. 
The simplest approach conceptually is called
{\it Gell-Mann-Low renormalization}.  This is the approach that was 
followed in the original development of renormalized
perturbation theory for Quantum Electrodynamics.
A quantum field theory model can be specified by a Hamiltonian 
density with a finite number of interaction terms:
\begin{eqnarray}
{\mathcal H} = {\mathcal H}_{\rm free} 
+ \sum_{n=1}^N c_n(\Lambda) {\mathcal O}_n .
\label{H:GML}
\end{eqnarray}
If the coupling constants $c_n(\Lambda)$ can be tuned as functions
of the ultraviolet cutoff so that physical observables 
have well-behaved and nontrivial limits as $\Lambda \to \infty$, 
the model is called {\it renormalizable}.
It may be necessary to add operators to the set
$\{ {\mathcal O}_1, \ldots, {\mathcal O}_N \}$ 
or remove some of the operators to make 
the model renormalizable.  In nonrelativistic quantum field theories 
in which the total number of particles cannot change,
the physical observables can be classified according to the particle 
number.  If the coupling constants can be tuned so that all 
$N$-particle observables have well-behaved limits as
$\Lambda \to \infty$, the model is called 
{\it renormalizable in the $N$-atom sector}.
If the model is renormalizable in the $N$-atom sector,
it is renormalizable in the $n$-atom sector for all $n < N$.
In this paper, we will for the most part consider only the constraints 
from renormalizability in the 2-atom sector.

The other approach to renormalization is called 
{\it Wilsonian renormalization}.  This approach was followed by 
Ken Wilson when he developed the renormalization group 
to understand strongly-interacting quantum field theories.
One begins by considering the most general model consistent 
with the symmetries.  The Hamiltonian density in Eq.~(\ref{H:GML})
is generalized to one with infinitely many operators ${\mathcal O}_n$,
each with its own coupling constant $c_n(\Lambda)$.
The free term ${\mathcal H}_{\rm free}$ is included as just another 
operator with a coupling constant $c_{\rm free}(\Lambda)$.
We define the engineering dimension $d_n$ of an operator
${\mathcal O}_n$ by specifying the dimensions of $\psi$ 
and $\nabla$ to be $\frac32$ and 1, respectively.
For this purpose, the mass $m$ is treated  
as a dimensionless conversion factor.
The engineering dimension of the free term in Eq.~(\ref{H:free})
is then 5.  An $N$-atom operator with $N$ factors 
of $\psi$, $N$ factors of $\psi^\dagger$,
and $D$ factors of the gradient $\nabla$ has engineering 
dimension $3 N + D$.  It is convenient to introduce 
dimensionless coupling constants $\hat c_n(\Lambda)$
defined by
\begin{eqnarray}
\hat c_n(\Lambda) = \Lambda^{d_n-5} c_n(\Lambda).
\end{eqnarray}
The condition that low-energy observables defined by the 
Hamiltonian be independent of $\Lambda$ defines a flow on the 
infinite-dimensional space of 
dimensionless coupling constants $\hat c_n$.
This flow can have a {\it fixed point} whose coordinates $c_n^*$ 
are independent of $\Lambda$.  The corresponding fixed-point 
Hamiltonian ${\mathcal H}^*$ describes a scale-invariant system.
Near the fixed point, a convenient basis of operators consists 
of {\it scaling operators} ${\mathcal O}_n^*$
that have the property that when their coupling constants are 
infinitesimally small, they scale as definite powers of $\Lambda$:
\begin{eqnarray}
\hat c_n(\Lambda)  \sim \Lambda^{\gamma_n}.  
\end{eqnarray}
We refer to the exponent $\gamma_n$ as the {\it scaling dimension}
of the operator ${\mathcal O}_n^*$.  
An operator with scaling dimension $\gamma_n<0$ 
is called {\it relevant}, because it becomes increasingly important as
$\Lambda$ decreases towards 0 and thus has a dramatic effect on 
the physics at low energies.  An operator with scaling dimension 
$\gamma_n>0$ is called {\it irrelevant}, because it decreases
in importance as $\Lambda$ decreases towards 0.
An operator with scaling dimension $\gamma_n=0$ is called {\it marginal}.
The Hamiltonian density can be expanded around the 
fixed point in terms of the scaling operators:
\begin{eqnarray}
{\mathcal H} = {\mathcal H}^* 
+ \sum_{n=1}^\infty c_n(\Lambda) {\mathcal O}_n^* .
\label{H:Wilson}
\end{eqnarray}
Once the scaling operators 
have been identified and their scaling dimensions $\gamma_n$ determined, 
the model can be truncated to include only a finite number of terms 
in the Hamiltonian.
A consistent truncation must include all the operators that are
relevant and marginal.
The behavior of the system near the fixed point can be described 
more accurately by including irrelevant operators.
The accuracy of the description can be systematically improved by adding
operators with increasing values of $\gamma_n$.

The simplest example of a renormalization group fixed point is the 
free theory with the Hamiltonian density ${\mathcal H}_{\rm free}$
in Eq.~(\ref{H:free}).  This fixed point is called the 
{\it trivial fixed point}.  The forms of the scaling operators
depend on the choice of ultraviolet cutoff or regularization.
With a scale-invariant regularization scheme,
such as dimensional regularization,
the scaling operators consist of terms that all have the 
same engineering dimensions.  
A scaling operator with $N$ factors of $\psi$, 
$N$ factors of $\psi^\dagger$,
and $D$ factors of $\nabla$ has scaling dimension 
$3 N + D -5$. The 2-body interaction terms 
in Eq.~(\ref{H:int}) are irrelevant operators with
scaling dimensions 1 and 3, respectively.
The simplest 3-body interaction term $(\psi^\dagger)^3 \psi^3$
is an irrelevant operator with scaling dimension 4.
The successive truncation of the Hamiltonian to include operators 
with increasingly higher scaling dimension corresponds to a 
combination of the gradient expansion and an expansion in the number 
of particles.

The system of nonrelativistic particles with short-range interactions
also has a nontrivial fixed point.  In terms of the effective range 
expansion in Eq.~(\ref{effrangeexp}), the fixed point corresponds 
to the limit in which the scattering length $a$ is infinite and 
the effective range $r_s$ and all higher coefficients in the 
expansion are 0.  The dimensions of the scaling operators 
at the fixed point were deduced by Kaplan, Savage and Wise
\cite{Kaplan:1998tg,Kaplan:1998we} and by van Kolck
\cite{vanKolck:1998bw}.  
There is a relevant operator with scaling dimension $-1$ 
which if added to the fixed-point Hamiltonian 
gives a model with a finite scattering length.   
There are also irrelevant operators 
with scaling dimensions 1, 3, 5, \ldots.

\section{Zero-Range Model}
\label{sec:zrm}

The simplest scattering model for ultracold atoms
is the {\it Zero-Range Model}, whose only parameter is
the scattering length $a$.
The Zero-Range Model can be defined by the truncation of the
effective range expansion in Eq.~(\ref{effrangeexp}) after the
constant term:
\begin{eqnarray}
k \cot \delta_0(k) = - 1/a.
\label{delta0:zrm}
\end{eqnarray}
The effective range and all the higher coefficients in the
effective range expansion in Eq.~(\ref{effrangeexp}) are exactly 0.

The Zero-Range Model provides a natural description for atoms
with a large scattering length that arises from the accidental
fine tuning of the depth of the interatomic potential.
In such a case, the scattering length typically changes very slowly 
with the magnetic field $B$.  Its dependence on $B$
can be taken into account by
allowing the parameter $a$ in Eq.~(\ref{delta0:zrm}) 
to be a smooth function of $B$.
The Zero-Range Model also provides a minimal description for atoms
with a large scattering length that arises from any other
mechanism.  If that mechanism is a Feshbach resonance,
the dependence of the scattering length $a$ on $B$
has the form given in Eq.~(\ref{a:FR}).

\subsection{Hamiltonian}

The Zero-Range Model can be described by a quantum field theory
with the single complex field $\psi$. 
The Hamiltonian density is the sum of the free term
in Eq.~(\ref{H:free})
and the interaction term
\begin{eqnarray}
{\mathcal H}_{\rm int} & = &
{1 \over 4} \lambda_0 (\psi^\dagger \psi)^2.
\label {H:zrm}
\end{eqnarray}
The interaction term corresponds to a momentum-independent
vertex with the Feynman rule $- i \lambda_0$.
Such an interaction is singular, so the model requires an
ultraviolet cutoff $\Lambda$ on the momenta of virtual atoms.

The Zero-Range Model is renormalizable in the 2-body sector.
As shown in Sec.~\ref{sec:zrem-Green},
the bare coupling constant $\lambda_0$
in Eq.~(\ref {H:zrm}) can be adjusted as a function of $\Lambda$
so that 2-body observables
do not depend on $\Lambda$.  The scattering length has the
desired value $a$ if the bare coupling constant is
\begin{eqnarray}
\lambda_0(\Lambda) =
\left ( 1 - {2 \over \pi} a \Lambda \right)^{-1} {8 \pi a \over m}.
\label{lambda0:contact}
\end{eqnarray}
This expression can be inverted to express the scattering length
as a function of the bare coupling constant and the cutoff:
\begin{eqnarray}
a = \frac{m}{8 \pi}
\left[ \frac{1}{\lambda_0} + {m \over 4 \pi^2}  \Lambda  \right]^{-1} .
\label{a-lambda0}
\end{eqnarray}

The expression for the bare coupling constant in 
Eq.~(\ref{lambda0:contact}) provides a simple illustration 
of Wilsonian renormalization, which was discussed in 
Sec.~\ref{sec:renorm}.
The operator $(\psi^\dagger \psi)^2$ in Eq.~(\ref{H:zrm})
has engineering dimension 6.
The corresponding dimensionless coupling constant is therefore
$\Lambda \lambda_0(\Lambda)$.
This dimensionless coupling constant has fixed points for 
two values of the scattering length: $a = 0$ and $a = \pm \infty$.
The trivial fixed point $a=0$ corresponds to a noninteracting theory.
For infinitesimal values of $a$, the dimensionless coupling constant
is $\Lambda \lambda_0(\Lambda) \approx 8 \pi a \Lambda/m$.
Thus near the trivial fixed point,
$(\psi^\dagger \psi)^2$ is an irrelevant operator 
with scaling dimension 1.
The nontrivial fixed point $a = \pm \infty$ is commonly 
referred to as the {\it unitary limit}.  At the fixed point, the
dimensionless coupling constant is
$\hat \lambda_0^* \approx - 4 \pi^2/m$.
For infinitesimal values of $1/a$, the deviation of the dimensionless 
coupling constant from its fixed-point value is
$\Lambda \lambda_0(\Lambda) - \hat \lambda_0^* 
\approx -2 \pi^3/(m a \Lambda)$.
Thus near the unitary fixed point,
$(\psi^\dagger \psi)^2$ is a relevant operator 
with scaling dimension $-1$.

\subsection{Green's function}
\label{sec:zrem-Green}

\begin{figure}[tb]
\centerline{\includegraphics*[width=12.8cm,angle=0,clip=true]{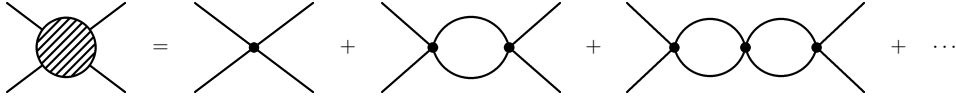}}
\caption{
The series of diagrams whose sum is the amputated connected
Green's function for atom-atom scattering in a model with
contact interactions only, such as the Zero-Range Model.}
\label{fig:2->2(zrm)}
\end{figure}

\begin{figure}[tb]
\centerline{\includegraphics*[width=8cm,angle=0,clip=true]{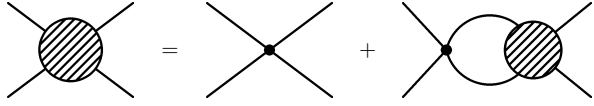}}
\caption{
The integral equation for the amputated connected
Green's function for atom-atom scattering in a model with
contact interactions only, such as the Zero-Range Model.}
\label{fig:inteq(zrm)}
\end{figure}

The amputated connected Green's function for 2 atoms
to evolve into 2 atoms can be calculated analytically
by summing the series of diagrams shown in Fig.~\ref{fig:2->2(zrm)}.
Each vertex gives a factor of $- i \lambda_0$.
Each loop gives the product of two propagators, which must be integrated 
over the momenta and energies of the virtual atoms.
In the center-of-mass frame, the integral is
\begin{eqnarray}
&&\int \frac{d\omega}{2 \pi} \int \frac{d^3k}{(2 \pi)^3} \,
\frac{i}{\omega - k^2/(2m) + i \varepsilon} \,
\frac{i}{E-\omega - k^2/(2m) + i \varepsilon} 
\nonumber
\\
&& \hspace{3cm}
= \int \frac{d^3k}{(2 \pi)^3} \frac{i}{E - k^2/m + i \varepsilon} ,
\label{A-loop}
\end{eqnarray}
where $E$ is the total energy in that frame.
The energy integral has been evaluated using contour integration.
The remaining momentum integral is ultraviolet divergent.
It can be regularized by imposing a momentum cutoff: $|\bm{k}|< \Lambda$.  
For each loop, there is also a symmetry factor of $\frac{1}{2}$. 
The resulting expression for the amplitude is
\begin{eqnarray}
i{\mathcal A} (E) = - i \lambda_0 \sum_{n=0}^\infty
\left[  (- i \lambda_0) \frac{1}{2}
\int \frac{d^3k}{(2 \pi)^3} \frac{i}{E - k^2/m + i \varepsilon} \right]^n .
\label{A-series}
\end{eqnarray}
The amplitude depends on the total energy $E$ of the two atoms
in the center-of-mass frame, but not on the momenta
of the incoming or outgoing atoms.
The sum of the geometric series in Eq.~(\ref{A-series}) is
\begin{eqnarray}
{\mathcal A} (E) =
- \lambda_0 \left( 1 - \frac{\lambda_0}{2} I_0(E) \right)^{-1}  ,
\label{A-sum}
\end{eqnarray}
where the integral $I_0(E)$ is defined in Eq.~(\ref{In-E}).
The expression for the integral
is particularly simple if we take
$\Lambda$ to be so much larger than $(m |E|)^{1/2}$ that we can
neglect terms that decrease as inverse powers of
$\Lambda/(m |E|)^{1/2}$ as $\Lambda\to \infty$.
The integral is given in the Appendix in Eq.~(\ref{int-0}).
The resulting expression for the Green's function is
\begin{eqnarray}
{\mathcal A} (E) = -
\left[ \frac{1}{\lambda_0} + {m \over 4 \pi^2} \Lambda
- {m \over 8 \pi} \kappa  \right]^{-1} ,
\label{A:bare}
\end{eqnarray}
where $\kappa$ is proportional to the square root of the energy:
\begin{eqnarray}
\kappa = (-mE - i \varepsilon)^{1/2}.
\label{kappa-E}
\end{eqnarray}
The amplitude in Eq.~(\ref{A:bare}) can also be obtained by
solving the Lippmann-Schwinger integral equation, which is represented 
diagrammatically in Fig.~\ref{fig:inteq(zrm)} and can be written
\begin{eqnarray}
i{\mathcal A} (E) = -i \lambda_0 + (-i \lambda_0)
\frac{1}{2} \int \frac{d^3k}{(2 \pi)^3} \frac{i}{E - k^2/m + i \varepsilon}
\big( i{\mathcal A} (E) \big).
\end{eqnarray}
Since ${\mathcal A} (E)$ does not depend on the momenta
of the incoming or outgoing particles, this is just a linear
equation for ${\mathcal A} (E)$
and its solution is Eq.~(\ref{A:bare}).

The dependence of the amplitude in Eq.~(\ref{A:bare})
on the ultraviolet cutoff can be eliminated by choosing 
the bare coupling constant $\lambda_0(\Lambda)$ to have the form in 
Eq.~(\ref{lambda0:contact}), where $a$ is an arbitrary parameter 
with dimensions of length.
The renormalized expression for the amplitude is then
obtained by using Eq.~(\ref{lambda0:contact})
to eliminate $\lambda_0$ in favor of $a$:
\begin{eqnarray}
{\mathcal A} (E) = - {8 \pi  \over m} \,
{a \over 1 - a \kappa}.
\label{A:zrm}
\end{eqnarray}
Since ${\mathcal A}(E)$ encodes all the physical 2-body observables,
we conclude that the Zero-Range Model is renormalizable in the
2-body sector.

A simple way to reproduce the renormalized expression for
${\mathcal A}(E)$ in Eq.~(\ref{A:zrm}) is to use 
{\it dimensional regularization},
which automatically subtracts power ultraviolet divergences
such as the linear term in $\Lambda$ in Eq.~(\ref{int-0}).
With this regularization scheme, the integral $I_0(E)$
is given simply by Eq.~(\ref{I0dr}).  Since no renormalization
is necessary in this case, we call the parameter $\lambda$
instead of $\lambda_0$.  The amplitude in
Eq.~(\ref{A-sum}) then reduces to
\begin{eqnarray}
{\mathcal A} (E) = -
\left[ \frac{1}{\lambda} - {m \over 8 \pi}  \kappa  \right]^{-1} .
\label{A:baredimreg}
\end{eqnarray}
This agrees with Eq.~(\ref{A:zrm}) if we set
$\lambda =8 \pi a/m$.

\subsection{T-matrix element}

The T-matrix element for atom-atom scattering 
with relative momentum $k$ can be obtained from
${\mathcal A}(E)$ in Eq.~(\ref{A:zrm}) by setting $E$ equal to the
total kinetic energy $2 (k^2 / 2m)$ of the two atoms:
\begin{eqnarray}
{\mathcal T}(k) = - {8 \pi  \over m} \, {a \over 1 + i ak}.
\label{T:zrm}
\end{eqnarray}
By taking the low-energy limit $k \to 0$ and comparing with 
Eq.~(\ref{Tfm}), we identify $a$ as the scattering length.
Comparing with Eq.~(\ref{T-delta}), we obtain the simple equation
for the phase shift in Eq.~(\ref{delta0:zrm}).

\subsection{Bound state}

The amplitude ${\mathcal A}(E)$ in Eq.~(\ref{A:zrm}) has a pole
in the energy variable $\kappa$ defined in Eq.~(\ref{kappa-E}).
The pole is at $\kappa = 1/a$.
If $a > 0$, this corresponds to a pole in the energy
at $E = - 1/(ma^2)$, which implies that there is a stable bound state
with binding energy $1/(m a^2)$.
This bound state is the {\it shallow dimer}, whose binding energy 
is given in Eq.~(\ref{E-dimer}).
If $a<0$, the pole at $\kappa = 1/a$ corresponds to pole 
at $E = - 1/(ma^2)$ on the second sheet of the complex variable $E$.
Such a pole corresponds to a virtual state.

\subsection{Optical Theorem}

If the energy $E$ is real, the imaginary part of the amplitude 
in Eq.~(\ref{A:zrm}) is
\begin{eqnarray}
{\rm Im} \, {\mathcal A} (E) = 
{(8 \pi/m) a^2k \over 1 + a^2 k^2} \theta(E)
+ {16 \pi^2/m^2  \over a} \, 
\delta(E + 1/(m a^2)) \, \theta(a) .
\label{optical:zrm}
\end{eqnarray}
The first term is nonzero for positive values of $E = k^2/m$.  
The second term, which contributes only if $a>0$,
is a delta function at the negative energy $-E_D$.
Note that the expression for ${\rm Im} \, {\mathcal A} (E)$
in Eq.~(\ref{optical:zrm}) is positive definite,
in accord with the Optical Theorem in Eq.~(\ref{optical}).
This implies that there are no negative-norm states 
in the 2-atom sector of the Zero-Range Model.

\subsection{Three-body sector}

In the 2-body sector, the Zero-Range Model is completely defined
by a single parameter: the scattering length $a$.
In the 3-body and higher sectors,
this condition is not sufficient to specify the model completely
because of the {\it Efimov effect} \cite{Efimov70,Efimov71}.
In the unitary limit $a \to \pm \infty$, there are infinitely-many
arbitrarily-shallow trimers called {\it Efimov states}
whose energies $-E_T^{(n)}$ have an accumulation point at the 3-atom threshold.
For $a = \pm \infty$, the spectrum is
\begin{eqnarray}
E_T^{(n)} =
 (e^{2 \pi/s_0})^{n_*-n} {\hbar^2 \kappa_*^2 \over m} ,
\label{ET}
\end{eqnarray}
where $n_*$ is an integer, $\kappa_*$ has dimensions of momentum,
and $s_0$ is the transcendental number that satisfies
\begin{eqnarray}
s_0 \cosh \frac{\pi s_0}{2} = \frac{8}{\sqrt{3}} \sinh \frac{\pi s_0}{6}.
\label{s0}
\end{eqnarray}
Its numerical value is  $s_0 \approx 1.00624$.
The spectrum in Eq.~(\ref{ET}) defines a 3-body parameter 
$\kappa_*$ that is independent of $a$.
One can interpret $\hbar^2 \kappa_*^2/m$
as the approximate binding energy of the trimer labeled $n_*$.
A different choice of $n_*$ gives a value of $\kappa_*$
that differs by an integer power of $e^{\pi/s_0} \approx 22.7$.
The ratio of successive binding energies in Eq.~(\ref{ET})
is $e^{2 \pi/s_0} \approx 515.03$.
This spectrum in Eq.~(\ref{ET}) is consistent with an asymptotic 
discrete scaling symmetry with discrete scaling factor 
$e^{\pi/s_0} \approx 22.7$. 
The 3-body observables are uniquely determined
only if one specifies both the scattering length $a$
and the 3-body parameter $\kappa_*$.
The 3-body parameter $\kappa_*$ can be adjusted by adding a 
$(\psi^\dagger \psi)^3$ term to the Hamiltonian  
\cite{Bedaque:1998kg,Bedaque:1998km}.
The renormalization of its coefficient is governed by a 
limit cycle instead of a fixed point \cite{Braaten:2004rn}.
Thus if the Zero-Range Model is extended to include a 
$(\psi^\dagger \psi)^3$ interaction term, it is renormalizable 
in the 3-atom sector.

It is an open question whether the Zero-Range Model is renormalizable 
in the 4-atom and higher sectors.  One problem  is that the spectrum 
of Efimov states in Eq.~(\ref{ET}) is unbounded from below in the 
limit $\Lambda \to \infty$, where $\Lambda$ is the ultraviolet cutoff.
Thus there can be no stable 4-atom bound states in the 
Zero-Range Model, because they can decay into an Efimov state 
and an atom.  In spite of this pathology of the model, it is still 
possible that low-energy observables in the 4-atom and higher 
sectors have well-defined universal limits as $\Lambda \to \infty$
that depend only on the parameters $a$ and $\kappa_*$.
There is numerical evidence that this is the case
in the 4-atom sector \cite{Platter:2004qn,Hammer:2006ct}.


\section{Effective Range Model}
\label{sec:erm}

The simplest generalization of the Zero-Range Model is the
{\it Effective Range Model}, whose two parameters are
the scattering length $a$ and the effective range $r_s$.
The Effective Range Model can be defined by the truncation of the
effective range expansion in Eq.~(\ref{effrangeexp}) after the
$k^2$ term:
\begin{eqnarray}
k \cot \delta_0(k) = - 1/a + \mbox{$1\over2$} r_s k^2 .
\label{delta0:ert}
\end{eqnarray}
All the higher coefficients in the effective range expansion are 0.
The Effective Range Model provides a natural description for  
atoms that have a large scattering length and a large effective range 
because of a double fine-tuning.
It can also describe atoms with a large scattering length and  
a natural effective range more accurately than the Zero-Range Model.

Phillips, Beane, and Cohen showed that the phase shift in
Eq.~(\ref{delta0:ert}) can be reproduced
by a renormalizable local Hamiltonian quantum field theory 
with a single atom field only if the effective range $r_s$ 
is negative \cite{Phillips:1997xu}.  This will be verified below 
in Sec.~\ref{sec:ermT}.
In Sec.~\ref{sec:rm}, we will show that
the phase shift in Eq.~(\ref{delta0:ert}) can also be reproduced by
a renormalizable local Hamiltonian quantum field theory with
an atom field and a molecular field, but again only if $r_s < 0$.
In Sec.~\ref{sec:otherTD}, we will show that the phase shift in 
Eq.~(\ref{delta0:ert}) with either sign of $r_s$ can be reproduced 
by a Lagrangian quantum field theory with a 
time-derivative interaction.
However, if $r_s > 0$, the price that must be paid is a space of
quantum states that includes negative-norm states.

\subsection{Hamiltonian}
\label{sec:Herm}

The simplest Hamiltonian quantum field theory 
that gives a nonzero effective range  has the 
two interaction terms in Eq.~(\ref{H:int}),
which can also be expressed in the form
\begin{eqnarray}
{\mathcal H}_{\rm int} & = &
{1 \over 4} \lambda_0 (\psi^\dagger \psi)^2
+ {1 \over 4} \rho_0 \nabla (\psi^\dagger \psi) \cdot \nabla (\psi^\dagger \psi).
\label{Hint-mer}
\end{eqnarray}
The second interaction term can be expressed in an 
alternative form using Eq.~(\ref{nabla^2}):
\begin{eqnarray}
{\mathcal H}_{\rm int} =
\frac{1}{4} \lambda_0 (\psi^2)^\dagger \psi^2
- \frac{1}{16} \rho_0
\left[ (\psi^2)^\dagger
(\psi {\stackrel \leftrightarrow \nabla}{}^2 \psi) +
(\psi {\stackrel \leftrightarrow \nabla}{}^2 \psi)^\dagger \psi^2 \right].
\label{Hint-mer:2}
\end{eqnarray}
The difference between the Hamiltonian
densities in Eqs.~(\ref{Hint-mer}) and (\ref{Hint-mer:2}) is
a total divergence, so it gives zero when integrated over all space
to get the Hamiltonian.
The elementary vertex for particles with momenta $\bm{p}_1$ and
$\bm{p}_2$ to scatter into momenta $\bm{p}_1'$ and $\bm{p}_2'$ is
\begin{eqnarray}
-i \left( \lambda_0  + \frac{1}{4} \rho_0
\left[ (\bm{p}_1 - \bm{p}_2)^2 + (\bm{p}_1' - \bm{p}_2')^2 \right] \right)
= -i \left[ \lambda_0 + \rho_0 (k^2 + {k'}^2) \right] ,
\end{eqnarray}
where $k={1\over 2}|\bm{p}_1 - \bm{p}_2|$
and $k'={1\over 2}|\bm{p}_1' - \bm{p}_2'|$
are the relative momenta in the initial and final states.

As shown by Phillips, Beane, and Cohen,
the Effective Range Model is renormalizable in the 2-body sector
provided the effective range $r_s$ is negative \cite{Phillips:1997xu}.
The bare coupling constants $\lambda_0$ and $\rho_0$
in Eq.~(\ref{Hint-mer:2}) can be adjusted as functions of 
the ultraviolet cutoff $\Lambda$
so that 2-body Green's functions and 2-body observables
have finite limits as $\Lambda \to \infty$.  
The intricate tuning that is required will be derived in 
Sec.~\ref{sec:ermT}:
\begin{subequations}
\begin{eqnarray}
\rho_0(\Lambda) &\longrightarrow& - {12 \pi^2 \over m \Lambda^3}
\left[ 1 \pm \left( {-12 \over \pi r_s \Lambda} \right)^{1/2}
\left( 1 - {\pi \over 2 a \Lambda} + {6 \over \pi r_s \Lambda} + \ldots \right) \right] ,
\label{rho0:ert}
\\
\lambda_0(\Lambda) &\longrightarrow& {m \over 20 \pi^2} \rho_0(\Lambda)^2 \Lambda^5
+ {48 \pi \over m r_s \Lambda^2} \,
\left( 1 - {\pi \over 2 a \Lambda} + {12 \over \pi r_s \Lambda} + \ldots \right) .
\label{lam0:ert}
\end{eqnarray}
\label{rholam:ert}
\end{subequations}
Note that the dimensionless coupling constants
$\Lambda \lambda_0$ and $\Lambda^3 \rho_0$ 
approach finite limits as $\Lambda \to \infty$.

\subsection{Green's function}

The amputated connected Green's function for two atoms
can be calculated analytically by solving the Lippmann-Schwinger integral
equation represented diagramatically in Fig.~\ref{fig:inteq(zrm)}.
This equation can be solved most easily by assuming that the
amplitude can be expressed in the form
\begin{eqnarray}
{\mathcal A} (E;k,k') &=&  {\mathcal A}_0(E)
+ {\mathcal A}_1(E) \big( k^2 + {k'}^2 \big)
+ {\mathcal A}_2(E) k^2 {k'}^2 .
\end{eqnarray}
The integral equation then reduces to a system
of linear equations for the three functions
${\mathcal A}_0(E)$, ${\mathcal A}_1(E)$, and ${\mathcal A}_2(E)$
\cite{Phillips:1997xu}.
The final result for the amplitude can be expressed in the form
\begin{eqnarray}
{\mathcal A} (E;k,k') &=& - \frac{N(E;k,k')}{D(E)} .
\label{A:merm0}
\end{eqnarray}
The numerator is a function of $E$, $k$, $k'$, and $\Lambda$:
\begin{eqnarray}
N(E;k,k') &=&
\left( \lambda_0 - \frac{m}{20 \pi^2} \rho_0^2 \Lambda^5 \right)
- 2 \rho_0 \left( 1 + \frac{m}{24 \pi^2} \rho_0 \Lambda^3 \right) \kappa^2
\nonumber
\\
&&+ \rho_0 \left( 1 + \frac{m}{12 \pi^2} \rho_0 \Lambda^3 \right)
    \left(2 \kappa^2 + k^2 + {k'}^2 \right)
\nonumber
\\
&& - \frac{m}{4 \pi^2} \rho_0^2\left( \Lambda - \frac{\pi}{2} \kappa \right)
    \left(\kappa^2 + k^2\right) \left(\kappa^2 + {k'}^2 \right) ,
\label{N:merm}
\end{eqnarray}
where $\kappa= (-mE - i \varepsilon)^{1/2}$.
The denominator is a function of $E$ and $\Lambda$:
\begin{eqnarray}
D(E) &=&
\left( 1 + \frac{m}{12 \pi^2} \rho_0 \Lambda^3 \right)^2
+ \frac{m}{4 \pi^2}
\Bigg[ \left( \lambda_0 - \frac{m}{20 \pi^2} \rho_0^2 \Lambda^5 \right)
\nonumber
\\
&&   \hspace{4cm} 
- 2 \rho_0 \left( 1 + \frac{m}{24 \pi^2} \rho_0 \Lambda^3 \right)
    \kappa^2 \Bigg]
    \left( \Lambda - \frac{\pi}{2} \kappa \right) .
\label{D:merm}
\end{eqnarray}

Since the numerator is a linear combination of four independent
functions of $\kappa$, $k$, and $k'$, the dependence of the amplitude
${\mathcal A} (E;k,k')$ on $\Lambda$ cannot be completely eliminated
simply by taking the two bare parameters $\lambda_0$ and $\rho_0$
to be functions of $\Lambda$.  However, as shown by Phillips, Beane, 
and Cohen \cite{Phillips:1997xu}, if the bare parameters
$\lambda_0$ and $\rho_0$ are tuned as in Eqs.~(\ref{rholam:ert}), 
the ratio in Eq.~(\ref{A:merm0}) will have a finite
limit as $\Lambda \to \infty$.  The renormalized amplitude 
obtained by this limiting procedure has the form
\begin{eqnarray}
{\mathcal A} (E) &=& - {8 \pi \over m} \,
{a \over 1 + {1\over2} r_s a \kappa^2 - a \kappa} .
\label{A:ert}
\end{eqnarray}
It depends on the total energy $E$ 
but not on the momenta $k$ and $k'$.

\subsection{T-matrix element}
\label{sec:ermT}

The T-matrix element for atom-atom scattering 
with relative momentum $k$ can be obtained from
${\mathcal A}(E)$ in Eq.~(\ref{A:ert}) by setting $E$
equal to the total energy $k^2/m$ of the two atoms:
\begin{eqnarray}
{\mathcal T} (k) &=& - {8 \pi \over m} \,
{a \over 1 - {1\over2} r_s a k^2 + i a k} .
\label{T:ert}
\end{eqnarray}
Comparing with Eq.~(\ref{T-delta}), we obtain the equation
for the phase shift in Eq.~(\ref{delta0:ert}).

To drive the expressions for the tuning of the bare parameters
in Eqs.~(\ref{rholam:ert}), we start from the unrenormalized expression for the 
T-matrix element obtained by setting $k' = k$ and $\kappa = -i k$
in the  Green's function
${\mathcal A} (E;k,k')$ given by Eqs.~(\ref{A:merm0}), (\ref{N:merm}), and
(\ref{D:merm}):
\begin{eqnarray}
{\mathcal T}(k) &=& - \Bigg(
{[1 + (m/12 \pi^2) \rho_0 \Lambda^3]^2
\over [\lambda_0 - (m/20 \pi^2) \rho_0^2 \Lambda^5]
    + 2 \rho_0 [1 + (m/24 \pi^2) \rho_0 \Lambda^3] k^2}
\nonumber
\\
&& \hspace{7cm}
+ {m \over 4 \pi^2}\Lambda + i {m \over 8\pi} k \Bigg)^{-1}.
\label{Tert}
\end{eqnarray}
Expanding ${\mathcal T}(k)^{-1}$ in powers of $k$ and comparing with
the effective range expansion in Eq.~(\ref{effrangeexp}), we find that the
scattering length and the effective range are
\begin{subequations}
\begin{eqnarray}
a &=& \frac{m}{8\pi}
\frac{1}
{[1 + (m/12 \pi^2) \rho_0 \Lambda^3]^2/
    [\lambda_0 - (m/20 \pi^2) \rho_0^2 \Lambda^5]
+ (m/4 \pi^2) \Lambda},
\label{a-ert}
\\
r_s &=& \frac{32\pi} {m}
\frac{ \rho_0 [1 + (m/12 \pi^2) \rho_0 \Lambda^3]^2
    [1 + (m/24 \pi^2) \rho_0 \Lambda^3]}
{[\lambda_0 - (m/20 \pi^2) \rho_0^2 \Lambda^5]^2} .
\label{rs-ert}
\end{eqnarray}
\end{subequations}
Using Eq.~(\ref{a-ert}) to eliminate $\lambda_0$ in favor of $a$,
the expression for the effective range in Eq.~(\ref{rs-ert}) becomes
\begin{eqnarray}
r_s =
\frac{2m}{\pi^3}
\frac{\rho_0 [1 + (m/24 \pi^2) \rho_0 \Lambda^3]}
{[1 + (m/12 \pi^2) \rho_0 \Lambda^3]^2} \,
\left[ \Lambda - \frac{\pi}{2a} \right]^2.
\label{rs-ert:2}
\end{eqnarray}
One can show that for the right side of Eq.~(\ref{rs-ert:2})
to have a finite nonzero limit as $\Lambda \to \infty$,
the quantity $1 + (m/12 \pi^2) \rho_0 \Lambda^3$
in the denominator must approach 0 as $\Lambda^{-1/2}$:
\begin{eqnarray}
 [1 + (m/12 \pi^2) \rho_0 \Lambda^3]^2
\longrightarrow
- \frac{12}{\pi r_s \Lambda}
\left( 1 - {\pi \over a \Lambda} + {12 \over \pi r_s \Lambda} + \ldots \right) .
\label{rho0-lim}
\end{eqnarray}
This equation implies that the bare parameter $\rho_0$
must have the limiting behavior given in Eq.~(\ref{rho0:ert}).

If we insert the limiting expression in Eq.~(\ref{rho0-lim}) into the
expression for the scattering length in Eq.~(\ref{a-ert}),
it reduces to
\begin{eqnarray}
a = \frac{m}{8\pi}
\Bigg[ {-12/(\pi r_s \Lambda)
    \over \lambda_0 - (m/20 \pi^2) \rho_0^2 \Lambda^5}
\left( 1 - {\pi \over a \Lambda} + {12 \over \pi r_s \Lambda} + \ldots \right)
+ {m \over 4 \pi^2} \Lambda \Bigg]^{-1} .
\end{eqnarray}
This shows that the quantity $\lambda_0 - (m/20 \pi^2) \rho_0^2 \Lambda^5$
must approach 0 as $\Lambda^{-2}$:
\begin{eqnarray}
\lambda_0 - {m \over 20 \pi^2} \rho_0^2 \Lambda^5
\longrightarrow
 {48 \pi \over m r_s \Lambda^2}
\left( 1 - {\pi \over 2 a \Lambda} + {12 \over \pi r_s \Lambda} + \ldots \right).
\label{lambda0-lim}
\end{eqnarray}
It also implies that the bare parameter $\lambda_0$
must have the limiting behavior given in Eq.~(\ref{lam0:ert}).

After inserting the limiting expressions in Eqs.~(\ref{rho0-lim})
and (\ref{lambda0-lim}), the T-matrix element in Eq.~(\ref{Tert})
reduces to
\begin{eqnarray}
{\mathcal T} (k) =
- \frac{8\pi}{m} \lim_{\Lambda \to \infty}
\left[
- {2 \over \pi} \Lambda
\left( 1 + {\pi \over 2 a \Lambda} - {\pi r_s \over 4 \Lambda} k^2 \right)^{-1}
+ {2 \over \pi} \Lambda + i k \right]^{-1} .
\label{T-erm:2}
\end{eqnarray}
Taking the limit $\Lambda \to \infty$, we recover the
renormalized expression for the T-matrix element in Eq.~(\ref{T:ert}).

Since the left side of Eq.~(\ref{rho0-lim}) is positive definite,
this equation implies that the
effective range $r_s$ in this model must be negative.
This result was first derived by Phillips, Beane, and Cohen
\cite{Phillips:1997xu}.  They showed that this conclusion
is more general.  The effective range
must be negative for any model with
a Hamiltonian that includes gradient interactions of arbitrarily high
order.  This result is consistent with a proof that the
self-adjoint extension of the Hamiltonian with a delta-function
potential to a 2-parameter Hamiltonian with scattering length
$a$ and effective range $r_s$ can be made on a Hilbert
space with positive definite norm only if $r_s <0$ \cite{Fewster:1994sd}.

\subsection{Bound states}

The amplitude ${\mathcal A}(E)$ in Eq.~(\ref{A:ert}) has two poles 
in the energy variable $\kappa$ defined by Eq.~(\ref{kappa-E}).
The values of $\kappa$ at the poles satisfy a quadratic equation:
\begin{eqnarray}
1 - a \kappa +\mbox{$1\over2$} r_s a \kappa^2 = 0.
\label{kappaD:ert}
\end{eqnarray}
If there is a positive real root $\kappa_i$, there is a stable bound state
with binding energy $\kappa_i^2/m$. If there is a negative real root  
$\kappa_i$, the pole corresponds to a virtual state.
The quadratic polynomial in Eq.~(\ref{kappaD:ert}) 
has no real roots if $a/r_s < 2$. 
If $a/r_s > 2$, it has two real roots:
\begin{eqnarray}
\kappa_\pm = {1 \over r_s} \left(1 \pm \sqrt{1-2 r_s/a} \right) .
\end{eqnarray}
The condition $a/r_s > 2$ requires $a$ and $r_s$ to 
either have opposite signs or to have the same signs 
and satisfy $|a| > 2 |r_s|$.
The root $\kappa_-$ is positive if $a>0$ and negative if $a<0$.  
The root $\kappa_+$ is positive if $r_s>0$ and negative if $r_s<0$.
Thus there can be 0, 1, or 2 stable bound states depending on the signs 
of $a$ and $r_s$.

If $|a|$ is large compared to $|r_s|$, the approximate solutions 
are $\kappa_- \approx 1/a$ and $\kappa_+ \approx 2/r_s$.  
The solution $\kappa_-$
corresponds to the shallow dimer if $a>0$ and to a shallow
virtual state if $a<0$.  If $r_s < 0$, as required by the 
renormalizability of the Effective Range Model,
the solution $\kappa_+$ corresponds to a virtual state.  
If $r_s > 0$ was allowed, the solution 
$\kappa_+$ would correspond to a stable deeply-bound state.

\subsection{Optical theorem}

If the energy $E$ is real and $a/r_s > 2$, 
the imaginary part of the amplitude 
in Eq.~(\ref{A:ert}) is
\begin{eqnarray}
{\rm Im} \, {\mathcal A} (E) &=& 
{(8 \pi/m)a^2 k \over (1 - {1\over2} r_s a k^2)^2 + a^2 k^2} \theta(E)
\nonumber
\\
&& 
\hspace{-0.2cm}
+ {16 \pi^2/m^2 \over \sqrt{1 - 2 r_s/a}} 
\big[ \kappa_- \, \delta(E + \kappa_-^2/m) \, \theta(a)
- \kappa_+ \, \delta(E + \kappa_+^2/m) \, \theta(r_s) \big].
\nonumber
\\
\end{eqnarray}
The first term is nonzero for positive values of $E = k^2/m$.  
This is the only term if $a/r_s < 2$.  If $a/r_s > 2$,
there can also be delta function contributions
at negative values of $E$.
There is a delta function term  with a positive coefficient 
if $a > 0$ and another delta function term with a 
negative coefficient if $r_s > 0$.
Thus if $r_s > 0$, the Optical Theorem includes a 
negative-definite term as in Eq.~(\ref{optical:+-}).
The bound state associated with this term can be interpreted 
as a state with negative norm.  The condition $r_s < 0$ for the 
absence of negative-norm states coincides with the condition 
for the renormalizability of the Effective Range Model.
If there is a large scattering length $|a| \gg r_s$,
the negative-norm state corresponds to a deeply-bound state.


\section{Other Single-Channel models}
\label{sec:oscm}

In this sector, we describe briefly other scattering models
that can be formulated as local quantum field theories
with a single quantum field $\psi$.

\subsection{Higher-gradient interactions}

One straightforward generalization of the Effective Range Model 
is to include higher gradient terms
in the interaction Hamiltonian.
There are two independent terms with four gradients
that give only $S$-wave scattering:
\begin{eqnarray}
\Delta {\mathcal H}_{\rm int} & = &
\mu_0 
\left(
\psi {\stackrel \leftrightarrow \nabla}{}^2 \psi
\right)^ \dagger
\left(
\psi {\stackrel \leftrightarrow \nabla}{}^2 \psi
\right)
\nonumber
\\
& + &
\mu_0 ' 
\left[
(\psi ^2)^ \dagger
\left( \psi
{\stackrel \leftrightarrow \nabla}{}^2
{\stackrel \leftrightarrow \nabla}{}^2
\psi \right)
+
\left( \psi
{\stackrel \leftrightarrow \nabla}{}^2
{\stackrel \leftrightarrow \nabla}{}^2
\psi \right)^ \dagger
\psi ^2
\right] .
\label{Hint-new}
\end{eqnarray}
The model whose Hamiltonian is the sum of
${\mathcal H}_{\rm free}$ in Eq.~(\ref{H:free}), 
${\mathcal H}_{\rm int}$ in Eq.~(\ref{Hint-mer}), 
and $\Delta {\mathcal H}_{\rm int}$ in  Eq.~(\ref{Hint-new})
has four bare parameters
that can depend on the ultraviolet cutoff.
If this model is renormalizable,
one might expect it to be able to reproduce additional terms
in the effective range expansion of
the $S$-wave phase shift in Eq.~(\ref{effrangeexp})
beyond the two terms that are shown.
Such a model is not of much practical use for atoms,
because the $1/r^6$ tail of the van der Waals potential 
implies that $k \cot \delta_0(k)$ has an expansion 
in powers of $k^2$ only to order $k^2$.

In Ref.~\cite{Yang:2004ss}, the Green's function
${\mathcal A} (E; k,k')$
for this model with an ultraviolet cutoff $\Lambda$
was obtained by solving the Lippmann-Schwinger integral
equation shown diagrammatically in Fig.~\ref{fig:inteq(zrm)}.
The authors showed that the dependence of
the T-matrix element on the ultraviolet cutoff
could not be removed simply by choosing
the bare parameters to be functions of $\Lambda$,
as in the Zero-Range Model in Sec.~\ref{sec:zrm}.
However, they did not investigate whether
the model is renormalizable in the sense of
having a nontrivial limit as $\Lambda \to \infty$,
as in the Effective Range Model in Sec.~\ref{sec:erm}.
Since the Effective Range Model is the special case 
$\mu_0 = \mu_0'=0$, it is clear that
the four bare parameters can be chosen as functions of $\Lambda$
so that the phase shift in the limit $\Lambda \to \infty$
is Eq.~(\ref{delta0:ert}) with $r_s < 0$.
It is not known whether a more general phase shift
can be obtained by taking such a limit.

\subsection{Dimensional regularization}

If there is a consistent way
to renormalize a quantum field theory,
one can often obtain the final result more simply by
using dimensional regularization.
In this regularization method, ultraviolet-divergent integrals 
over a $3$-dimensional momentum are 
generalized to a variable number of
dimension $d$, evaluated in a range of $d$ for which
they converge, and then analytically continued to $d=3$.
This regularization prescription treats
power divergences very differently
from logarithmic divergences.
Power-divergent integrals are those
that diverge as a positive power $\Lambda^p$
with a momentum cutoff $\Lambda$.
Logarithmically-divergent integrals are those
that diverge as $\log \Lambda$ as $\Lambda \to \infty$.
Dimensional regularization sets power divergence to $0$,
while logarithmic divergences appear as poles
in $d-3$.

In the Effective Range Model, which is defined by the interaction Hamiltonian
in Eq.~(\ref{Hint-mer}), the only ultraviolet divergences
that appear are power divergences $\Lambda^p$ with $p=1,3,5$.
Dimensional regularization sets these divergences to $0$.
Thus the Green's function ${\mathcal A} (E;k,k')$
with dimensional regularization can be obtained
from the results in
Eqs.~(\ref{A:merm0}), (\ref{N:merm}), and (\ref{D:merm})
with a momentum cutoff simply by setting $\Lambda=0$.
The resulting expression is
\begin{eqnarray}
{\mathcal A} (E;k,k') &=& -
\frac{\lambda
    + \rho ( k^2 + {k'}^2)
    +  (m/8 \pi) \rho^2 \kappa(\kappa^2 + k^2) (\kappa^2 + {k'}^2)}
{1 -  (m/8 \pi)(\lambda - 2 \rho \kappa^2) \kappa} .
\label{A-ermdr}
\end{eqnarray}
Since no renormalization is required,
we have replaced the bare parameters
$\lambda_0$ and $\rho_0$ by $\lambda$ and $\rho$.
The T-matrix element ${\mathcal T}(k)$ for atom-atom scattering
can be obtained from the amplitude
${\mathcal A} (E;k,k')$ in Eq.~(\ref{A-ermdr})
by setting $k' = k$ and $\kappa = -i k$:
\begin{eqnarray}
{\mathcal T} (k) &=& - \,
{1 \over 1/(\lambda + 2 \rho k^2) + i (m/8\pi)k} .
\label{T-ermdr:1}
\end{eqnarray}
Expanding ${\mathcal T}(k)^{-1}$ in powers of $k$ and comparing with
the effective range expansion in Eq.~(\ref{effrangeexp}), we find that
the parameters $\lambda$ and $\rho$ are related in a simple way to the
scattering length and the effective range:
$\lambda = 8 \pi a/m$, $\rho = 2 \pi a^2 r_s/m$.
If $\lambda$ and $\rho$ are eliminated in favor of $a$ and $r_s$,
the T-matrix element in Eq.~(\ref{T-ermdr:1}) reduces to
\begin{eqnarray}
{\mathcal T} (k) &=& - {8 \pi \over m} \,
{a \over 1/(1 + {1\over2} r_s a k^2) + i a k} .
\label{T-ermdr:2}
\end{eqnarray}
This result corresponds to the $S$-wave phase shift
\begin{eqnarray}
k \cot \delta_0(k) = - {1 \over a \left( 1 + {1\over2} r_s a k^2 \right)}.
\label{delta0:merm}
\end{eqnarray}

The phase shift in Eq.~(\ref{delta0:merm})
obtained using dimensional regularization
has a different dependence on $k$ from
the phase shift in Eq.~(\ref{delta0:ert})
obtained using a momentum cutoff regularization.
Renormalization of the Effective Range Model using a momentum
cutoff requires the constraint $r_s < 0$.
There is no apparent constraint
on $r_s$ if we use dimensional regularization.
This puzzle was first pointed out in Ref.~\cite{Phillips:1997xu}.
The fact that dimensional regularization of the Effective Range Model gives
the phase shift in Eq.~(\ref{delta0:merm})
suggests that there might be a more elaborate model with
a momentum cutoff that gives this phase shift.
However, the results of Ref.~\cite{Phillips:1997xu}
imply that such a model cannot be a 
Hamiltonian quantum field theory.

\subsection{Time-derivative interactions}
\label{sec:otherTD}

Interactions that involve time derivatives
are not allowed in a conventional
Hamiltonian quantum field theory. However, they can
be easily included in the Lagrangian approach.
We will consider only the simplest Lagrangian model with
a time-derivative interaction and show that
it is equivalent to the Effective Range Model.

The Lagrangian density for a free atom is
given in Eq.~(\ref{Lfree}).
The interaction term in the Lagrangian density for
the Zero-Range Model is ${\mathcal L}_{\rm int} = -{\mathcal H}_{\rm int}$,
where ${\mathcal H}_{\rm int}$ is given in Eq.~(\ref{H:zrm}).  
The simplest generalization of the Zero-Range Model
that includes a time-derivative interaction
is a model whose interaction term is
\begin{eqnarray}
{\mathcal L}_{\rm int} = - \frac{1}{4} (\psi^2)^\dagger
\left(
\lambda_0 - m \lambda_0 '
    \left(
    i \frac{\partial} {\partial t} + \frac{\nabla^2}{4m}
    \right)
\right) \psi^2 .
\label{Lint:time}
\end{eqnarray}
The combination of time and space derivatives
acting on $\psi ^2$ is Galilean invariant.

The diagrammatic analysis of the model
whose Lagrangian is the sum of
Eqs.~(\ref{Lfree}) and (\ref{Lint:time})
is identical to that in the Zero-Range Model
except that the interaction vertex depends on
the total energy $E$ of the interacting atoms
in their center-of-mass frame.
The amplitude in this model can be obtained from the
unrenromalized Green's function
for the Zero-Range Model in Eq.~(\ref{A:bare})
simply by the substitution
$\lambda_0 \to \lambda_0 + \lambda_0 ' \kappa ^2$,
where $\kappa = (-mE -i \epsilon)^{1/2}$.
The resulting expression for the Green's function is
\begin{eqnarray}
{\mathcal A} (E) = - \left[
\frac{1}{\lambda_0 + \lambda_0 ' \kappa ^2}
+
{m \over 4 \pi^2} \Lambda
- {m \over 8 \pi} \kappa
\right]^{-1} .
\label{A:bare-new}
\end{eqnarray}
We obtain the $T$-matrix element ${\mathcal T} (k)$
by setting $\kappa = -i k$.
By using the expression for ${\mathcal T} (k)$ in
Eq.~(\ref{T-delta}) and the expression for
$k \cot \delta_0 (k)$ in Eq.~(\ref{effrangeexp}),
we can determine the scattering length and effective range:
\begin{subequations}
\begin{eqnarray}
\frac{1}{a} &=& \frac{8 \pi}{m}
\left( \frac{1}{\lambda_0} +
\frac{m}{4 \pi^2} \Lambda \right),
\\
r_s &=& - \frac{16 \pi}{m}
\frac{\lambda_0 '} {\lambda_0 ^2}.
\end{eqnarray}
\end{subequations}
By inverting these equations, we can determine
how the bare parameters $\lambda_0$ and $\lambda_0 '$
must be tuned as functions of
the cutoff $\Lambda$ in order to
keep $a$ and $r_s$ fixed:
\begin{subequations}
\begin{eqnarray}
\lambda_0(\Lambda) &=&
\left ( 1 - \frac{2}{\pi} a \Lambda
\right)^{-1}
{8 \pi a \over m},
\\
\lambda_0 ' (\Lambda) &=&
- r_s \frac{4 \pi a^2}{m}
\left ( 1 - \frac{2}{\pi} a \Lambda
\right)^{-2}.
\label{lambdaprime:bare}
\end{eqnarray}
\end{subequations}
Inserting these expressions into Eq.~(\ref{A:bare-new})
and taking the limit $\Lambda \to \infty$,
we obtain the renormalized amplitude:
\begin{eqnarray}
{\mathcal A} (E) = - \frac{8 \pi} {m} \,
\left[
{1 \over a} + {1 \over 2} r_s \kappa^2 - \kappa
\right]^{-1}.
\end{eqnarray}
This is identical to the renormalized amplitude
for the Effective Range Model in Eq.~(\ref{A:ert}).
Since all 2-body observables can be obtained
from the Green's functions, we conclude that 
the Lagrangian field theory with the time-derivative interaction 
in Eq.~(\ref{Lint:time}) is equivalent to the Effective Range Model,
at least in the 2-body sector.

\subsection{Wilsonian renormalization}

The Wilsonian approach to renormalization 
requires finding a fixed-point Hamiltonian ${\mathcal H}^*$
together with scaling operators ${\mathcal O}_n^*$ 
and their scaling dimensions $\gamma_n$.
The Hamiltonian can then be expanded around the fixed point 
as in Eq.~(\ref{H:Wilson}).  The Wilsonian approach can also be 
applied to the larger space of Lagrangian quantum field theories.
In this case, there is a similar expansion around a fixed-point 
Lagrangian ${\mathcal L}^*$.

In a remarkable paper, Birse, McGovern, and Richardson
carried out a complete Wilsonian analysis of 2-body S-wave 
interactions \cite{Birse:1998dk}.
In a Lagrangian quantum field theory, the general S-wave 
2-body operator can be expressed in terms of a potential 
$V(p, k,k',\Lambda)$ that depends on the total energy
$E = p^2/m$, the squares $k^2$ and $k^{'2}$ of the initial 
and final momenta, and the ultraviolet cutoff $\Lambda$.
The requirement that the solution ${\mathcal A}(p, k,k')$
to the Lippmann-Schwinger integral equation be 
independent of the ultraviolet cutoff $\Lambda$ gives
renormalization group flow equations on the 
infinite-dimensional space of potentials.
The trivial fixed point is $V=0$.  There is also a 
nontrivial fixed point corresponding to 
the unitary limit of infinite scattering length:%
\footnote{This differs from the expression in Ref.~\cite{Birse:1998dk}
by a factor of 2 associated with identical particles.} 
\begin{eqnarray}
V^*(p,\Lambda) = - \frac{4 \pi^2} {m} \,
\left[ \Lambda 
- \frac{p}{2} \log \frac{\Lambda + p}{\Lambda - p} \right]^{-1} .
\label{V-fixed}
\end{eqnarray}
This can be expanded as a power series in $p^2$,
so it corresponds to an interaction term with arbitrarily 
high-order time derivatives.
A system infinitesimally close to the fixed point has
a potential of the form $V^*(p,\Lambda)+ v(p, k,k',\Lambda)$.
The authors of Ref.~\cite{Birse:1998dk} found all the 
scaling operators $v(p, k,k',\Lambda)$.  
The scaling operators $v(p,\Lambda)$ that depend only on the
energy variable $p$ have the scaling dimensions $-1$, 1, 3, 5, 
\ldots\ that were deduced in 
Refs.~\cite{Kaplan:1998tg,Kaplan:1998we,vanKolck:1998bw}.
There are also scaling operators $v(p, k,k',\Lambda)$
that have nontrivial dependence on the momentum variables $k$ and $k'$.
They have scaling dimensions 2, 4, 5, 6, $\ldots$.
They may correspond to redundant operators in the Lagrangian 
formulation that have no counterparts in a Hamiltonian formulation
and do not affect physical observables.

Harada and Kubo and collaborators have applied the Wilsonian 
renormalization approach to Lagrangian field theories that 
are truncated in the derivative expansion
\cite{Harada:2005tw,Harada:2006cw,Harada:2007ua}.
In Refs.~\cite{Harada:2005tw,Harada:2006cw}
they considered the 3-parameter Lagrangian consisting of 
the free term in Eq.~(\ref{Lfree}), the two interaction terms 
of the Effective Range Model in Eq.~(\ref{Hint-mer:2}), 
and the time-derivative interaction in Eq.~(\ref{Lint:time}).
This corresponds to a potential of the form
\begin{eqnarray}
V(p,k, k',\Lambda) = \lambda_0(\Lambda)
+ \rho_0(\Lambda) (k^2 + k^{'2} )
- \lambda_0'(\Lambda) p^2 .
\label{V-trunc}
\end{eqnarray}
By demanding that the solution to the Lippmann-Schwinger 
integral equation have a cutoff-independent expansion to 
first order in $p^2$, $k^2$, and $k^{'2}$, they obtained 
renormalization group flow equations for the coupling constants 
$\lambda_0(\Lambda)$, $\rho_0(\Lambda)$, and $\lambda_0'$.
These equations have 3 fixed points: the trivial fixed point,
the nontrivial fixed point, and an unphysical fixed point 
that gives complex scaling dimensions.
Near the nontrivial fixed point, there is a relevant 
operator with scaling dimension $-1$ and two irrelevant
operators with scaling dimensions 1 and 2. 
In Ref.~\cite{Harada:2007ua}, the analysis was extended 
to the next order in the derivative expansion, where there 
are 8 coupling constants.  They found three fixed points, two
fixed lines, and a fixed surface.  One of the fixed points 
was identified as the nontrivial fixed point 
associated with infinite scattering length.

\section{Two-Channel Model}
\label{sec:2cm}

A scattering model can describe multiple spin states of an atom.  
A minimal model has two spin states,
including the one of primary interest.
In general, there are 3 scattering 
channels corresponding to both atoms in the lower state, 
both atoms in the upper state, and atoms in both states.
We consider the simple case
of momentum-independent interactions
between the atoms, as in the Zero-Range Model.
We further simplify the problem by assuming that there is
nontrivial scattering only in the two channels in which both atoms 
are in the same spin state. 
We will refer to this model as the {\it Two-Channel Model}.
If the energies of atoms at rest in the two spin states 
are 0 and $\nu/2$, the scattering thresholds are 0 and $\nu$.
We choose $\nu>0$ and we take the spin state 
of primary interest to be the one that is lower in energy. 
The parameters of the Two-Channel Model can be defined by 
specifying the S-wave phase shift 
in the lower spin channel to be
\begin{eqnarray}
k \cot \delta_0(k) =
- {1 \over a_{11}}
- {1 \over a_{12}^2}
\Big[ -1/a_{22} + \sqrt{m \nu - k^2} \, \Big]^{-1} .
\label{kcot-2cm}
\end{eqnarray}
In addition to the energy gap $\nu$, there are three 
interaction parameters with dimensions of length:
$a_{11}$, $a_{12}$, and $a_{22}$.
The interaction parameters are defined in such a way that
the two channels decouple in the limit $a_{12} \to \pm \infty$.
In this limit, $a_{11}$ and $a_{22}$ reduce to the scattering
lengths for the two independent channels.

The scattering length  and the effective range 
for the lower spin channel are
\begin{subequations}
\begin{eqnarray}
a &=& \left( \frac{1}{a_{11}}
+ \frac{1}{a_{12}^2} \Big[ \sqrt{m \nu} - 1/a_{22} \, \Big]^{-1}
\right)^{-1},
\label{a-2cm}
\\
r_s &=& - {1 \over a_{12}^2 \sqrt{m\nu}}
 \Big[  \sqrt{m\nu} -1/a_{22}  \Big]^{-2}.
\label{rs-2cm}
\end{eqnarray}
\end{subequations}
Note that the effective range $r_s$ is negative definite.
There are various ways to tune the parameters to get a large
scattering length $a$.
If $a_{11} < a_{12}^2/a_{22}$, a large
scattering length can be obtained by tuning the energy gap $\nu$
to near the critical value 
$(1/a_{22} - a_{11} / a_{12}^2)^2/m$.
At $\nu$ approaches the critical value, the effective range 
approaches 
$r_s \to a_{22} a_{12}^4/[a_{11}^2 (a_{11} a_{22} - a_{12}^2)]$.  
A large scattering length can also be obtained
by tuning the interaction parameter $a_{11}$ 
to the critical value $- a_{12}^2 [ \sqrt{m \nu} - 1/a_{22} ]$.

A renormalizable Hamiltonian quantum field theory for
a Two-Channel Model with two  distinguishable particles 
with unequal masses was constructed
by Cohen, Gelman, and van Kolck \cite{Cohen:2004kf}.
An essentially equivalent model has been used to 
describe the effects of $\Delta \Delta$ states 
on the two-nucleon system \cite{Savage:1996tb}.
In Sec.~\ref{sec:Ncm}, we generalize the Two-Channel Model 
to one with $N$ scattering channels.

\subsection{Hamiltonian}

The Two-Channel Model can be formulated as a quantum field theory
with two complex fields $\psi$ and $\psi_2$.
The Hamiltonian density for the Two-Channel Model is
the sum of a free term and an interaction term:
\begin{subequations}
\begin{eqnarray}
{\mathcal H}_{\rm free} & = &
{1 \over 2 m}\nabla \psi^\dagger \cdot \nabla \psi
+ {1 \over 2 m}\nabla \psi_2^\dagger \cdot \nabla \psi_2
+ \frac{1}{2} \nu \psi_2^\dagger \psi_2 ,
\label{Hfree:2cm}
\\
{\mathcal H}_{\rm int} & = &
{1 \over 4} \lambda_{0,11} (\psi^\dagger \psi)^2
+ {1 \over 4} \lambda_{0,12}
\left[ (\psi^\dagger \psi_2)^2 + (\psi_2^\dagger \psi)^2  \right]
+ {1 \over 4} \lambda_{0,22} (\psi_2^\dagger \psi_2)^2.
\label{H:2cm}
\end{eqnarray}
\label{LH:2cm}
\end{subequations}
The Two-Channel Model is renormalizable in the 2-atom sector.
The relations between the physical parameters $a_{ij}$
and bare parameters $\lambda_{0,ij}$ are
\begin{subequations}
\begin{eqnarray}
{1 \over  a_{11}}
& = &
{8\pi \over m}
{\lambda_{0,22} \over \lambda_{0,11} \lambda_{0,22} - \lambda_{0,12}^2 }
+ {2\over \pi} \Lambda,
   \\
{1 \over  a_{12} }
& = &
- {8\pi \over m}
{\lambda_{0,12} \over \lambda_{0,11} \lambda_{0,22} - \lambda_{0,12}^2 },
\\
{1 \over  a_{22}}
& = &
{8\pi \over m}
{\lambda_{0,11} \over \lambda_{0,11} \lambda_{0,22} - \lambda_{0,12}^2 }
+ {2 \over \pi} \Lambda.
\end{eqnarray}
\label{a-lambda0:2cm}
\end{subequations}
These equations can be expressed more compactly in the form
\begin{eqnarray}
{1 \over a_{mn}} &=&
{8\pi \over m} \left( \lambda_0^{-1} \right)_{mn}
+ {2\over \pi} \Lambda \delta_{mn} ,
\label{renorm-2ch}
\end{eqnarray}
where $\lambda_0^{-1}$ is the inverse of the matrix $\lambda_0$
of coefficients $\lambda_{0,ij}$.

Renormalization invariants are functions of the bare parameters
that are equal to the same functions of the corresponding
renormalized parameters independent of the value of the ultraviolet cutoff.
Parameters that require no renormalization,
such as the energy gap $\nu$, are renormalization invariants.
An alternative choice for the remaining parameters
of the Two-Channel Model are the entries $\lambda_{mn}$ 
of the matrix $\lambda$ whose inverse has the entries
\begin{eqnarray}
\left( \lambda^{-1} \right)_{mn} &=&
{m \over 8\pi a_{mn}}  .
\end{eqnarray}
With this choice of renormalized parameters, the traceless
components of the matrices $\lambda^{-1}$ and $\lambda_0^{-1}$
are renormalization invariants:
\begin{subequations}
\begin{eqnarray}
\left( \lambda^{-1} \right)_{11} -
\left( \lambda^{-1} \right)_{22} &=&
\left( \lambda^{-1}_0 \right)_{11} -
\left( \lambda^{-1}_0 \right)_{22} ,
\\
\left( \lambda^{-1} \right)_{12} &=&
\left( \lambda^{-1}_0 \right)_{12} .
\end{eqnarray}
\end{subequations}
If we take the three independent interaction parameters 
to be these two renormalization invariants and 
${\rm tr}(\lambda_0^{-1})$, the only bare parameter that
must depend on the ultraviolet cutoff is 
${\rm tr}(\lambda_0^{-1})$.

\subsection{Green's function}

The amputated connected Green's function ${\mathcal A}(E)$
for the Two-Channel Model
is a $2\times 2$ matrix that depends on the total energy $E$
in the center-of-mass frame.
The inverse of the matrix ${\mathcal A}(E)$ can be calculated 
analytically by solving the coupled system of integral equations
represented diagramatically in Fig.~\ref{fig:inteq(zrm)}:
\begin{eqnarray}
 A(E)^{-1} =
\begin{pmatrix}
-(\lambda_0^{-1})_{11} -(m/4\pi^2)(\Lambda  -\pi \kappa/2)
~~~~~~~~~~~~~~~~~~ -(\lambda_0^{-1})_{12}   \\
-(\lambda_0^{-1})_{12} ~~~~~~~~ 
-(\lambda_0^{-1})_{22}
- (m/4\pi^2)(\Lambda - \pi \sqrt{m \nu +\kappa^2}/2)
\end{pmatrix},
\nonumber
\\
\end{eqnarray}
where $\kappa=(-mE-i\varepsilon)^{1/2}$.
Inserting the relations in Eq.~(\ref{renorm-2ch})
between the physical parameters $a_{ij}$ and the bare 
parameters $\lambda_{0,ij}$, the dependence on the ultraviolet
cutoff $\Lambda$ disappears and the inverse matrix reduces to
\begin{eqnarray}
{\mathcal A}(E)^{-1}= \frac{m}{8\pi}
\begin{pmatrix}
-1/a_{11} + \kappa   &  -1/a_{12}   \\
-1/a_{12} & -1/a_{22}  + \sqrt{m \nu +\kappa^2 }
\end{pmatrix}.
\label{Ainv-2cm}
\end{eqnarray}
The square roots are defined for negative real arguments by the
$i \varepsilon$ prescription in the definition of $\kappa$ in
Eq.~(\ref{kappa-E}).
The explicit expression for the entries
of this matrix are
\begin{subequations}
\begin{eqnarray}
{\mathcal A}_{11}(E) &=& {8 \pi \over m}
\left( - {1 \over a_{11}} + \kappa
- {1 \over a_{12}^2}
\Big[ -{1 \over a_{22}} + \sqrt{m \nu +\kappa^2 } \, \Big]^{-1}
\right)^{-1},
\label{A11-2cm}
\\
{\mathcal A}_{12}(E) &=& {8 \pi \over m}
\left( -{1 \over a_{12}}
+ a_{12} \left[ - {1 \over a_{11}} + \kappa \right]
    \left[ - {1 \over a_{22}} + \sqrt{m\nu + \kappa^2 } \right]
\right)^{-1},
\label{A12-2cm}
\\
{\mathcal A}_{22}(E) &=& {8 \pi \over m}
\left( - {1 \over a_{22}} + \sqrt{m \nu + \kappa^2 }
- {1 \over a_{12}^2}
\Big[ -{1 \over a_{11}} + \kappa \, \Big]^{-1}
\right)^{-1}.
\label{A22-2cm}
\end{eqnarray}
\label{A-2cm}
\end{subequations}
These expressions agree with the results of 
Ref.~\cite{Cohen:2004kf} except for a factor of two 
associated with identical particles.%
\footnote{There is an error in the component 
${\mathcal A}_{22}(E)$ in Ref.~\cite{Cohen:2004kf},
but it vanishes when the masses of the two particles 
are set equal.}

\subsection{T-matrix element}

The T-matrix element for the elastic scattering of atoms
in the first channel with relative momentum $k$ is obtained 
by evaluating ${\mathcal A}_{11}(E)$ in Eq.~(\ref{A11-2cm}) 
at the total energy $E = k^2/m$:
\begin{eqnarray}
{\mathcal T}_{11}(k) = {8 \pi \over m}
\left( - {1 \over a_{11}} - i k
- {1 \over a_{12}^2}
\Big[ - {1 \over a_{22}} + \sqrt{m \nu - k^2} \, \Big]^{-1}
\right)^{-1}.
\label{T11-2cm}
\end{eqnarray}
Comparing with Eq.~(\ref{T-delta}), we obtain the equation
for the phase shift in Eq.~(\ref{kcot-2cm}).

\subsection{Bound states}

The three entries of the matrix ${\mathcal A}(E)$ 
given by Eq.~(\ref{A-2cm}) all have poles at the same values
of the energy variable $\kappa$.
The values of $\kappa$ at the poles satisfy the equation
\begin{eqnarray}
\kappa = \frac{1}{a_{11}}
+ \frac{1}{a_{12}^2}
\Big[ - {1 \over a_{22}} +\sqrt{m \nu +  \kappa^2} \, \Big]^{-1}.
\label{kappa-2cm}
\end{eqnarray}
By solving for the square root and then squaring both sides 
of the equation, one can show that the roots of Eq.~(\ref{kappa-2cm})
are also roots of a quartic polynomial.
If the root $\kappa_i$ is real and positive, there is a stable bound state
below the scattering threshold for the first channel
with binding energy $\kappa_i^2/m$.
If $\sqrt{m \nu} > 1/a_{22}$, Eq.~(\ref{kappa-2cm})
can have 0 or 1 positive real roots.
If $a_{22} > 0$ and $\sqrt{m \nu} < 1/a_{22}$, 
Eq.~(\ref{kappa-2cm}) can have 1 or 2 positive real roots.
Thus the number of stable diatomic molecules 
can be 0, 1, or 2. 

If the scattering length $a$ is made large by tuning $\nu$ 
to near its critical value, Eq.~(\ref{kappa-2cm}) 
will have one small root $\kappa \approx 1/a$.  If $a>0$, 
the corresponding bound state is the shallow dimer. 
The remaining roots of  Eq.~(\ref{kappa-2cm}) satisfy
$|\kappa| \gg 1/|a|$.
As $a \to \pm \infty$,
these large roots approach solutions to a cubic equation:
\begin{eqnarray}
\kappa^3 - \frac{2}{a_{11}} \kappa^2
+ \left( \frac{1}{a_{11}^2} - \frac{2 a_{11}}{a_{12}^2 a_{22}} 
+ \frac{a_{11}^2}{a_{12}^4} \right) \kappa
+ \frac{2}{a_{12}^2} \left( \frac{1}{a_{22}} - \frac{a_{11}}{a_{12}^2} \right) 
\approx 0.
\label{kappa-2cm:large}
\end{eqnarray}
Since this cubic equation was derived by squaring a 
square root, some of its roots may not correspond to 
roots of Eq.~(\ref{kappa-2cm}).
If $a_{11} < 0$, the limiting form of Eq.~(\ref{kappa-2cm})  
has no large positive real roots.
If $a_{11} > 0$, it has one large positive real root.
Thus the number of stable deeply-bound diatomic molecules 
is either 0 or 1.

\subsection{Optical theorem}

If the energy $E$ is real, the imaginary part of the amplitude
in Eq.~(\ref{A11-2cm}) is
\begin{eqnarray}
{\rm Im} \, {\mathcal A}_{11} (E)
&=&
{8\pi \over m }
\left|  - {1 \over a_{11}} - i k
- {1 \over a_{12}^2}
\bigg[ -{1 \over a_{22}} + \sqrt{m \nu - k^2 - i \varepsilon} \bigg]^{-1} 
\, \right|^{-2}
\nonumber 
\\
&& \hspace{1cm} \times
\left( k 
+ {\sqrt{k^2-m\nu}/a_{12}^2 \over k^2 - m \nu + 1/a_{22}^2}  \theta(E - \nu) \right)
\, \theta(E)
\nonumber \\
&&
+ {16 \pi^2  \over m^2} \sum_i\kappa_i
\left[ 1
+ { a_{12}^2 (\kappa_i - 1/a_{11})^2 \kappa_i \over \sqrt{m\nu + \kappa_i^2} }
\right]^{-1}
\delta(E + \kappa_i^2/m) \, \theta(\kappa_i) ,
\nonumber
\\
\label{optical:2cm}
\end{eqnarray}
where the sum in the last term is over the 
0, 1, or 2 positive real roots $\kappa_i$ of Eq.~(\ref{kappa-2cm}).
The expression for ${\rm Im} \, {\mathcal A}_{11} (E)$
is positive definite,
in accord with the Optical Theorem in Eq.~(\ref{optical}).
This is consistent with the absence of negative-norm states 
in the 2-atom sector of the Two-Channel Model.

\subsection{Multi-channel models}
\label{sec:Ncm}

The Two-Channel Model 
can be generalized to one with $N$ spin states
with nontrivial scattering only between pairs of atoms 
in the same spin state.
We refer to this model as the {\it $N$-Channel model}.
It has ${1\over 2}(N^2 + 3N - 2)$ parameters:
the $N-1$ energy gaps $\nu_n$ between the lowest 2-atom threshold
($\nu_1 \equiv 0$) and the $n^{\rm th}$ 2-atom threshold and
${1\over 2}N(N+1)$ interaction parameters $a_{mn}$
with dimensions of length.
The parameters
in the $N$-Channel model can be defined by specifying the
S-wave phase shift to be
\begin{eqnarray}
 k \cot \delta_0(k) =
-{1\over a_{11} }  -  {1\over C_{11} (k^2) }
\sum_{n=2}^N  {C_{1n}(k^2)  \over  a_{1n} },
\label{phaseshift-Ncm}
\end{eqnarray}
where $C_{1n}(k^2)$ is the $1n$ cofactor of the $N\times N$
matrix whose $mn$ entry is
$(m/8\pi) (-1/a_{mn} +\sqrt{m\nu_n -k^2} \, \delta_{mn})$.
The scattering length and the effective range are
obtained from the low-momentum expansion
of Eq.~(\ref{phaseshift-Ncm}):
\begin{subequations}
\begin{eqnarray}
 a & = &
\left(  {1\over a_{11}}
       + {1 \over C_{11}(0) }
\sum_{n=2}^N  { C_{1n}(0)  \over  a_{1n} }
\right)^{-1} ,  \\
 r_s & = &  - 2
\sum_{n=2}^N
{C_{11}(0)  C_{1n}'(0)-C_{11}'(0) C_{1n}(0)
    \over  \, a_{1n} \, C_{11}^2(0)},
\end{eqnarray}
\end{subequations}
where
$C_{1n}'(k^2) = (d/dk^2) C_{1n}(k^2)$.

The $N$-Channel model can be formulated as a quantum field theory
with $N$ complex fields $\psi_{n}$, $n=1,\cdots,N$.
The Hamiltonian density for the $N$-Channel model is
the sum of a free term and an interaction term:
\begin{subequations}
\begin{eqnarray}
{\mathcal H}_{\rm free} & = &
\sum_{n=1}^{N} \left[
{1 \over 2 m}
\nabla \psi^\dagger_{n} \cdot \nabla \psi_{n}
+ {1\over 2} \nu_{n} \psi_{n}^\dagger \psi_{n}
\right],
\\
{\mathcal H}_{\rm int} &=&
{1 \over 4}
\sum_{m=1}^{N} \sum_{n=1}^{N}
   \lambda_{0,mn} (\psi_{m}^\dagger \psi_{n})^2.
\label{H:multi-cm}
\end{eqnarray}
\label{LH:multi-cm}
\end{subequations}
Note that $\nu_1 = 0$ by definition.
The $N$-Channel model is renormalizable in the 2-atom sector.
The energy gaps $\nu_n$ require no renormalization.
The relations between the physical interaction parameters $a_{mn}$
and the bare coupling constants $\lambda_{0,mn}(\Lambda)$ are
\begin{equation}
 {1 \over a_{mn}} = {8\pi \over m}
\left(\lambda_{0}^{-1} \right)_{mn}
+ {2\over \pi} \Lambda\, \delta_{mn}.
\label{renorm-Ncm}
\end{equation}

The amputated connected Green's function ${\mathcal A}(E)$
for this coupled-channel system is
a $N\times N$ matrix that depends on the total energy $E$ in the
center-of-mass frame. The inverse matrix ${\mathcal A}(E)^{-1}$
can be obtained analytically by solving the coupled-channel
integral equations 
represented diagramatically in Fig.~\ref{fig:inteq(zrm)}.
After the renormalization using Eq.~(\ref{renorm-Ncm}),
the $mn$ entry of the inverse matrix reduces to
\begin{eqnarray}
({\mathcal A}(E)^{-1})_{mn} = \frac{m}{8\pi}
\left[
-{1\over a_{mn}} + \sqrt{m\nu_m + \kappa^2 } \, \delta_{mn}
\right],
\label{A-Ncm-reverse}
\end{eqnarray}
where $\kappa = (-mE-i\varepsilon)^{1/2}$.
This inverse matrix can be inverted by using
the identity
\begin{eqnarray}
{\mathcal A}_{mn}(E) = 
\frac{C_{nm}(- \kappa^2)}{\det ({\mathcal A}(E)^{-1})},
\end{eqnarray}
where $C_{nm}(- \kappa^2)$ is the $nm$ cofactor
of the matrix ${\mathcal A}(E)^{-1}$
given in Eq.~(\ref{A-Ncm-reverse}).
Since the determinant is the sum of the products
of the entries in any row (or column) and their cofactors,
the $mn$ entry of the matrix ${\mathcal A}(E)$
reduces to
\begin{eqnarray}
{\mathcal A}_{mn} (E)  = \frac{8\pi}{m}
\left[
-  {1\over C_{nm}(mE) }
\sum_{l =1}^N
{ C_{nl}(mE) \over a_{nl}  }
+ \sqrt{m \nu_n  + \kappa^2 }
\, {C_{nn}(mE) \over C_{nm}(mE)}
\right]^{-1}.
\nonumber
\\
\label{A-Ncm}
\end{eqnarray}

The T-matrix element for elastic scattering in the first channel 
is obtained by evaluating the expression for ${\mathcal A}_{11} (E)$
given by Eq.~(\ref{A-Ncm}) at the energy $E=k^2/m$:
\begin{eqnarray}
 {\mathcal T}_{11} (k)
   = \frac{8\pi}{m}
\left[ - {1 \over a_{11} } - i k
-  {1\over C_{11} (k^2)}
\sum_{n = 2 }^N
{ C_{1n}(k^2) \over a_{1n}  }  \right]^{-1}.
\label{T-Ncm}
\end{eqnarray}
This T-matrix element gives
the S-wave phase shift for the $N$-Channel model
in Eq.~(\ref{phaseshift-Ncm}).

\section{Resonance Model}
\label{sec:rm}

The basic degrees of freedom in a scattering model can 
include molecular states as well as atoms.
A minimal model has a single diatomic molecule that
couples to a pair of atoms in the spin state of interest.
We consider the simplest case of a momentum-independent
interaction between the atoms and a
momentum-independent coupling of the molecule to a pair of atoms.
We treat the molecular degree of freedom as a point particle,
so that the only structure of the molecule arises from its coupling 
to the atoms.  We refer to this model  as the {\it Resonance Model}.
The parameters of the Resonance Model can be defined by specifying
the S-wave phase shift of the atoms to be
\begin{eqnarray}
k \cot \delta_0(k) = - {8 \pi \over m}
\left( \lambda - {m g^2 \over m \nu - k^2} \right)^{-1}.
\label{delta0:resonance}
\end{eqnarray}
The three parameters can be interpreted as
the detuning energy $\nu$ of the molecule relative to the 2-atom
threshold, the coupling strength $g$ of the molecule to a
pair of atoms, and the self-coupling strength $\lambda$ of the atoms.

The scattering length and the effective range are
\begin{subequations}
\begin{eqnarray}
a &=&  {m \over 8 \pi} \left( \lambda - {g^2 \over \nu} \right) ,
\label{a:resonance}
\\
r_s &=&  - 16 \pi \left( \lambda - {g^2 \over \nu} \right)^{-2}
{g^2 \over m^2 \nu^2}.
\label{r:resonance}
\end{eqnarray}
\end{subequations}
Note that the effective range is negative definite.
If $\lambda = 0$, the phase shift in
Eq.~(\ref{delta0:resonance}) has the same form as the phase shift 
for the Effective Range Model in Eq.~(\ref{delta0:ert}).
Thus the Effective Range Model with $r_s < 0$
is a special case of the Resonance Model.
The scattering length in Eq.~(\ref{a:resonance}) can be made large
by tuning the energy gap $\nu$ to near 0.  
The limiting value of the effective range is 
$r_s \to -16 \pi/(m^2 g^2)$.

The Resonance Model provides a natural description of atoms 
near a Feshbach resonance, where the dependence of the 
scattering length $a$ on the magnetic field $B$ has the form 
given in Eq.~(\ref{a:FR}).  That dependence can be reproduced 
by taking $\nu$ to be linear in the magnetic field $B$ 
while $g^2$ and $\lambda$ are proportional to $a_{\rm bg}$:
\begin{subequations}
\begin{eqnarray}
\nu  & = & -\mu_i \, (B - B_i),
\\
\lambda & = & {8 \pi a_{\rm bg} \over m},
\\
g^2   & = & -{8 \pi a_{\rm bg} \over m} \, \mu_i \, \Delta_i .
\end{eqnarray}
\label{coeff-phys:resonance}
\end{subequations}
The parameter $\mu_i$ in Eq.~(\ref{coeff-phys:resonance}) 
can be interpreted as the difference between the magnetic moment 
of the molecule and twice the magnetic moment of an isolated atom.
In general, both $a_{\rm bg}$ and $\mu_i$ 
can be slowly-varying functions of the magnetic field.

The Resonance Model was first introduced by Kaplan 
as a model for nonrelativistic particles with a large
scattering length \cite{Kaplan:1996nv}.  He used dimensional 
regularization which eliminated the need for explicit 
renormalization of the parameters.
The Resonance Model was constructed independently 
by Kokkelmans et al.~\cite{KMCWH02} as a model for atoms 
near a Feshbach resonance.  They derived the
renormalization of the parameters that is required to make 
the observables independent of the ultraviolet momentum cutoff.
In Sec.~\ref{sec:multi-resonance}, we generalize the Resonance Model 
to one in which the atoms are coupled to  
$N$ different molecular states.

\subsection{Hamiltonian}

The Resonance Model can be formulated as a quantum field theory
with two complex fields: $\psi$, which annihilates an atom,
and $\phi$, which annihilates a diatomic molecule.
The Hamiltonian density for the Resonance Model is the sum
of a free term and an interaction term:
\begin{subequations}
\begin{eqnarray}
{\mathcal H}_{\rm free} & = &
{1 \over 2m} \nabla \psi^\dagger \cdot \nabla \psi
+ {1 \over 4m} \nabla \phi^\dagger \cdot \nabla \phi
+ \nu_0 \phi^\dagger \phi ,
\\
{\mathcal H}_{\rm int} & = &
 {1 \over 2} g_0 \left(\phi^\dagger \psi^2 + \psi^{\dagger2} \phi \right)
+ {1 \over 4} \lambda_0 (\psi^\dagger \psi)^2.
\label{H:resonance}
\end{eqnarray}
\label{LH:resonance}
\end{subequations}
To avoid ultraviolet divergences, an ultraviolet cutoff $\Lambda$
must be imposed on the momenta of virtual particles.

The Resonance Model is renormalizable in the 2-atom sector, 
which consists of states containing two atoms or one diatomic molecule.
The values for the bare parameters that are required to
reproduce the phase shift in Eq.~(\ref{delta0:resonance})
are
\begin{subequations}
\begin{eqnarray}
\lambda_0 (\Lambda) & = & Z(\Lambda) \lambda,
\label{lambda0:resonance}
\\
g_0 (\Lambda) & = & Z(\Lambda) g ,
\label{g0:resonance}
\\
\nu_0 (\Lambda) & = & \nu - [1-Z(\Lambda)] g^2/\lambda,
\label{nu0:resonance}
\end{eqnarray}
\label{coeff0:resonance}
\end{subequations}
where the renormalization constant $Z$ is
\begin{eqnarray}
Z(\Lambda) = \left( 1 - {m \over 4 \pi^2} \lambda \Lambda \right)^{-1}.
\label{Z-Lambda}
\end{eqnarray}
Note that in the special case
$\lambda = 0$, which  corresponds to the Effective Range Model,
the only renormalization that is necessary is an additive 
renormalization of the detuning energy:
\begin{eqnarray}
\nu_0 (\Lambda) = \nu + {m \over 4 \pi^2} g^2 \Lambda.
\label{nu0:erm}
\end{eqnarray}
The bare parameter $g_0$ is equal to its renormalized
counterpart $g$.

Using the renormalizations of the parameters given in
Eqs.~(\ref{coeff0:resonance}), one can construct two
renormalization invariants:
\begin{subequations}
\begin{eqnarray}
g/\lambda & = & g_0/\lambda_0,
\label{g-lam:rgi}
\\
\nu - g^2/\lambda  & = & \nu_0 - g_0^2/\lambda_0.
\label{nu-g-lam:rgi}
\end{eqnarray}
\label{rgi:resonance}
\end{subequations}
If we take the three independent parameters to be
the two renormalization invariants in Eqs.~(\ref{rgi:resonance})
along with $\lambda_0$, the only parameter
that must be adjusted as a function of the 
ultraviolet cutoff is $\lambda_0(\Lambda)$.

\subsection{Green's function}

\begin{figure}[tb]
\centerline{\includegraphics*[width=12.8cm,angle=0,clip=true]{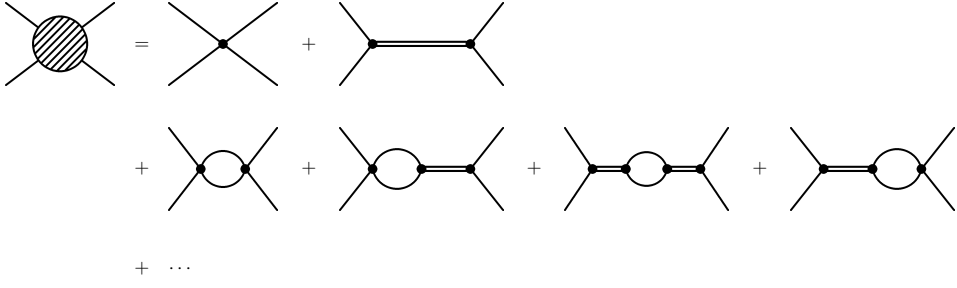}}
\caption{
The series of diagrams that give the amputated connected
Green's function for atom-atom scattering in the Resonance Model.}
\label{fig:2->2(RM)}
\end{figure}

\begin{figure}[tb]
\centerline{\includegraphics*[width=7.5cm,angle=0,clip=true]{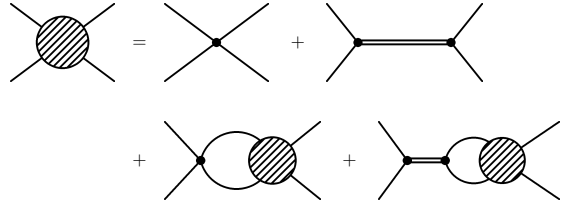}}
\caption{
The integral equation for the amputated connected
Green's function for atom-atom scattering in a model with
a diatomic molecule that has a coupling to two atoms,
such as the Resonance Model.}
\label{fig:inteq(RM)}
\end{figure}

The amputated connected Green's function 
for two atoms to evolve into two atoms
can be calculated analytically in the Resonance Model
by summing the series of diagrams shown in Fig.~\ref{fig:2->2(RM)}.
Alternatively, it can be obtained by solving the integral 
equation illustrated in Fig.~\ref{fig:inteq(RM)}.
This amplitude 
depends only on the total energy $E$ of the two atoms in
the center-of-mass frame:
\begin{eqnarray}
{\mathcal A} (E) = - \left[
\left( \lambda_0 + {g_0^2 \over E- \nu_0 + i \varepsilon} \right)^{-1}
+ {m \over 4 \pi^2} \Lambda - {m \over 8 \pi} \kappa \right]^{-1} ,
\end{eqnarray}
where $\kappa = (-mE - i \varepsilon)^{1/2}$.
After inserting the renormalization conditions 
in Eqs.~(\ref{coeff0:resonance}), the dependence on the ultraviolet
cutoff $\Lambda$ disappears and the amplitude reduces to
\begin{eqnarray}
{\mathcal A} (E) = - \left[
\left ( \lambda + {g^2 \over E- \nu + i \varepsilon} \right)^{-1}
- {m \over 8 \pi} \kappa \right ]^{-1} .
\label{A:resonance}
\end{eqnarray}
%

\subsection{T-matrix element}

The T-matrix element ${\mathcal T}(k)$ for the scattering 
of atoms with relative momentum $k$ can be obtained from
the amplitude ${\mathcal A} (E)$ in Eq.~(\ref{A:resonance}) 
by setting $E$ equal to the total energy $k^2/m$ 
of the two atoms:
\begin{eqnarray}
{\mathcal T} (k) =
- \left[ \left( \lambda - {mg^2 \over m \nu - k^2} \right)^{-1}
        + {i \over 8 \pi} m k \right ]^{-1}.
\label{T:resonance}
\end{eqnarray}
Comparing with Eq.~(\ref{T-delta}),
we obtain the S-wave phase shift given in Eq.~(\ref{delta0:resonance}).

\subsection{Bound states}

The amplitude ${\mathcal A} (E)$ in Eq.~(\ref{A:resonance}) 
has poles in the energy 
variable $\kappa$ defined by Eq.~(\ref{kappa-E}).
The values of $\kappa$ at the poles satisfy
\begin{eqnarray}
\left( \lambda - \frac{m g^2}{m \nu + \kappa^2} \right) ^{-1}
= \frac{m} {8 \pi} \kappa.
\label{pole-rm}
\end{eqnarray}
This can be expressed as a cubic polynomial equation.
The number of positive real roots is 0 if $\nu > 0$ and $\lambda < 0$,
1 if $\nu$ and $\lambda$ have the same sign,
and 2 if $\nu < 0$ and $\lambda > 0$. 
Thus the number of stable diatomic molecules can be 0, 1 or 2.

If the scattering length $a$ is made large by tuning $\nu$ 
to near 0, Eq.~(\ref{pole-rm}) will have one small root
$\kappa \approx 1/a$.  If $a>0$, the corresponding bound state is 
the shallow dimer.  The two remaining roots of 
Eq.~(\ref{pole-rm}) satisfy $|\kappa| \gg 1/|a|$.
As $a \to \pm \infty$,
these two large roots approach
\begin{eqnarray}
\kappa_\pm \approx 
\frac{4 \pi}{m \lambda}
\pm \left[ \left( \frac{4 \pi}{m \lambda} \right)^2 
+ \frac{mg^2}{\lambda} \right]^{1/2}.
\label{pole-rm:large}
\end{eqnarray}
If $\lambda < 0$, neither of these roots is real and positive.
If $\lambda > 0$, the root $\kappa_+$ is real and positive.
It corresponds to a stable deeply-bound diatomic molecule.

\subsection{Optical Theorem}

If the energy $E$ is real, the imaginary part of the amplitude
in Eq.~(\ref{A:resonance}) is
\begin{eqnarray}
{\rm Im} \, {\mathcal A} (E) &=&
\frac{m k}{8 \pi} \left[
\left( \lambda - {m g^2 \over m \nu - k^2} \right)^{-2}
+ \left( \frac{m k}{8 \pi} \right)^2
\right ]^{-1} \theta(E)
\nonumber \\
&+&
\frac{16 \pi^2} {m^2} \sum_i \kappa_i
\left[ 1 +
\frac{m^2 g^2 \kappa_i^3}{4 \pi (m \nu + \kappa_i^2)^2} \right]^{-1}
\delta(E + \kappa_i^2/m) \,  \theta(\kappa_i) ,
\label{optical:rm}
\end{eqnarray}
where the sum in the last term is over the 
0, 1, or 2 positive real roots $\kappa_i$ 
of Eq.~(\ref{pole-rm}).  The expression for
${\rm Im} \, {\mathcal A}(E)$ in Eq.~(\ref{optical:rm})
is positive definite, in accord with the Optical Theorem
in Eq.~(\ref{optical}).  This is consistent with the absence 
of negative-norm states in the 2-atom sector of the Resonance Model.

\subsection{Resonance Model with ghost molecule}

The expression in Eq.~(\ref{r:resonance}) for the effective range 
$r_s$ in the Resonance Model is negative definite.  
Thus this model cannot be used as
a phenomenological description of atoms with positive effective range.  
Kaplan pointed out that this limitation can be avoided by taking the
molecular field $\phi$ to be a 
{\it ghost field} whose quanta are states 
with negative norm \cite{Kaplan:1996nv}. 
The Lagrangian for this field theory differs from the Lagrangian 
for the Resonance Model only in the signs of a few terms.
The Lagrangian density is the sum of a free term and an 
interaction term:
\begin{eqnarray}
{\mathcal L}_{\rm free} & = &
\psi ^\dagger \left(
i \frac{\partial}{\partial t} + \frac{\nabla ^2} {2 m}
\right) \psi
+
\sigma \ \phi ^\dagger \left(
i \frac{\partial}{\partial t}
+ \frac{\nabla ^2} {4 m} - \sigma \ \nu_0
\right) \phi ,
\nonumber
\\
{\mathcal L}_{\rm int} & = &
- {1 \over 2} g_0 \left(
    \phi^\dagger \psi^2
    + \psi^{\dagger2} \phi
\right)
- {1 \over 4} \lambda_0 (\psi^\dagger \psi)^2,
\label{Leff-1}
\end{eqnarray}
where $\sigma = \pm 1$.
If $\sigma = + 1$, this is the Lagrangian density for the Resonance Model.
If $\sigma = - 1$, the free theory describes an atom and a ghost molecule.
In the Feynman rules for the interacting theory, the only difference 
from the Resonance Model is in the propagator for the molecule.
The propagator for a molecule with energy $E$ and momentum $\bm{P}$
is $i/[\sigma(E-P^2/(4m)) - \nu + i \varepsilon]$.
The expression for the Green's function in Eq.~(\ref{A:resonance})
is replaced by
\begin{eqnarray}
{\mathcal A} (E) = - \left[
\left ( \lambda + {g^2 \over \sigma \, E - \nu + i \varepsilon} \right)^{-1}
- {m \over 8 \pi} \kappa \right ]^{-1} .
\label{A:ghostresonance}
\end{eqnarray}
The S-wave phase shift in
Eq.~(\ref{delta0:resonance})
will be modified to
\begin{eqnarray}
k \cot \delta_0(k) =
- {8 \pi \over m}
\left( \lambda - {m g^2 \over
m \nu - \sigma \, k^2} \right)^{-1}.
\label{delta0:resonance1}
\end{eqnarray}
The scattering length is given by Eq.~(\ref{a:resonance})
for either sign $\sigma$.
The effective range differs from the expression in 
Eq.~(\ref{r:resonance}) by a factor of $\sigma$:
\begin{eqnarray}
r_s &=&  - 16 \pi \; \sigma
\left( \lambda - {g^2 \over \nu}
\right)^{-2}
{g^2 \over m^2 \nu^2}.
\label{r:resonance1}
\end{eqnarray}
If $\sigma = -1$, this is positive definite.

The amplitude in Eq.~(\ref{A:ghostresonance}) has poles in the energy 
variable $\kappa$ defined by Eq.~(\ref{kappa-E}).
The values of $\kappa$ at the poles are roots of the equation
\begin{eqnarray}
\left( \lambda - \frac{m g^2}{m \nu + \sigma \, \kappa^2} \right) ^{-1}
= \frac{m} {8 \pi} \kappa.
\label{pole-rmg}
\end{eqnarray}
If $\sigma = -1$,
the number of positive real roots can be 0 or 2 if 
$\nu$ and $\lambda$ have the same sign and 1 or 3 if 
they have opposite signs. 
If the energy $E$ is real, the imaginary part of the amplitude
in Eq.~(\ref{A:resonance}) is
\begin{eqnarray}
{\rm Im} \, {\mathcal A} (E) &=&
\frac{m k}{8 \pi} \left[
\left( \lambda - {m g^2 \over m \nu - \sigma \, k^2} \right)^{-2}
+ \left( \frac{m k}{8 \pi} \right)^2
\right ]^{-1} \theta(E)
\nonumber \\
&+&
\frac{16 \pi^2} {m^2} \sum_i \kappa_i
\left[ 1 +
\frac{\sigma \, m^2 g^2 \kappa_i^3}
    {4 \pi (m \nu + \sigma \, \kappa_i^2)^2} \right]^{-1}
\delta(E + \kappa_i^2/m) \,  \theta(\kappa_i) ,
\label{optical:rmg}
\end{eqnarray}
where the sum in the last term is over the 
0, 1, 2, or 3 positive real roots $\kappa_i$ of Eq.~(\ref{pole-rmg}).
Note that if $\sigma = -1$, the last term need not be 
positive definite.  Thus this model may have states 
that correspond to diatomic molecules with negative norm.

The scattering length in Eq.~(\ref{a:resonance}) can be made large 
by tuning the energy gap $\nu$ to be near 0.
The limiting value of the effective range for $\sigma = -1$ 
is $r_s \to 16 \pi/(m^2 g^2)$, which is positive.  
There is one small root of
Eq.~(\ref{pole-rmg}) that approaches $\kappa \approx 1/a$ in this limit.
If $a>0$, the corresponding bound state is the shallow dimer.  
For this root, the delta function contribution 
in Eq.~(\ref{optical:rmg}) has a positive coefficient,
so the shallow dimer has positive norm.
The two remaining roots of Eq.~(\ref{pole-rmg}) satisfy 
$|\kappa| \gg 1/|a|$.
As $a \to \pm \infty$, these two large roots approach
\begin{eqnarray}
\kappa_\pm \approx 
\frac{4 \pi}{m \lambda}
\pm \left[ \left( \frac{4 \pi}{m \lambda} \right)^2 
+ \sigma \frac{mg^2}{\lambda} \right]^{1/2}.
\label{kappa-rmg:large}
\end{eqnarray}
For the Resonance Model with a ghost molecule, we set $\sigma = -1$. 
A root $\kappa$ that is real and positive corresponds 
to a stable deeply-bound state. If the root is in the range 
$0 < \kappa < m^2 g^2/(4 \pi)$, the delta function 
contribution in Eq.~(\ref{optical:rmg}) has a negative coefficient,
so the bound state is a negative-norm state.
If $\lambda < 0$, the only large positive real root is $\kappa_+$,
and it corresponds to a negative-norm deeply-bound bound state. 
If $0< \lambda < 16 \pi^2/(m^3g^2)$, 
both $\kappa_+$ and $\kappa_-$ are positive real roots.
They correspond to a positive-norm and a negative-norm 
deeply-bound state, respectively.
If $\lambda > 16 \pi^2/(m^3g^2)$, there are no large positive real roots, 
so there are no stable deeply-bound states.
Even if it includes deeply-bound negative-norm states, the Resonance Model 
with $\sigma = -1$ may still be useful as an approximate 
description of atoms with a large scattering length and 
a positive effective range at energies small compared 
to $1/(m r_s^2)$.

\subsection{Multi-resonance models}
\label{sec:multi-resonance}

The {\it Resonance Model} can be
generalized to one with $N$ diatomic molecules that couple 
to a pair of atoms but not to each other.  
We refer to this model as the {\it $N$-Resonance Model}. 
It has $2N+1$ parameters:
the self-coupling strength $\lambda$ for the atoms,
a detuning energy $\nu_n$ for each of the $N$ molecules,
and a coupling strength $g_n$ for each of the $N$ molecules.
The parameters in the $N$-Resonance Model can be defined
by specifying the S-wave phase shift to be
\begin{eqnarray}
k \cot \delta_0(k) = - \frac{8 \pi} {m}
\left( \lambda -
\sum _{n=1} ^{N} \frac{m g_n ^2} {m \nu_n - k^2} \right)^{-1}.
\label{delta0:N-resonance}
\end{eqnarray}
The scattering length and the effective range are
\begin{subequations}
\begin{eqnarray}
a &=&  {m \over 8 \pi}
\left( \lambda
- \sum _{n=1} ^{N} \frac{g_n ^2} {\nu _n}
\right) ,
\label{a:N-resonance}
\\
r_s &=&  - 16 \pi
\left( \lambda
- \sum _{n=1} ^{N} \frac{g_n ^2} {\nu _n}
\right) ^{-2}
\sum _{n=1} ^{N}
\frac{g_n ^2} {m^2 \nu_n ^2} .
\label{r:N-resonance}
\end{eqnarray}
\end{subequations}
Note that the effective range is negative definite.

The Hamiltonian density for the $N$-Resonance Model is
the sum of a free term and an interaction term:
\begin{subequations}
\begin{eqnarray}
{\mathcal H}_{\rm free} & = &
\frac{1} {2m} \nabla \psi^\dagger \cdot \nabla \psi
+ \sum _{n=1} ^{N}
\left(
\frac{1} {4m} \nabla \phi_n ^\dagger \cdot \nabla \phi_n
+ \nu_{0,n} \ \phi_n ^\dagger \phi_n
\right) ,
\label{Hfree:N-resonance}
\\
{\mathcal H}_{\rm int} & = &
\frac{\lambda_0} {4} (\psi^\dagger \psi)^2
+ \sum _{n=1} ^{N} \frac{ g_{0,n} } {2}
\left(\phi_n ^\dagger \psi^2
+ \psi^{\dagger2} \phi_n \right) .
\label{Hint:N-resonance}
\end{eqnarray}
\label{LH:N-resonance}
\end{subequations}
As we will show below,
the relation between the bare coupling constants in Eq.~(\ref{LH:N-resonance})
and the renormalized parameters in Eq.~(\ref{delta0:N-resonance})
can be expressed compactly as an equality between two functions
of the energy $E$:
\begin{eqnarray}
\left( \lambda + \sum_{n=1}^N \frac{g_{n}^2}{E - \nu_{n}} \right)^{-1}
=
\left( \lambda_0
    + \sum_{n=1}^N \frac{g_{0,n}^2}{E - \nu_{0,n}} \right)^{-1}
+ \frac{m} {4 \pi^2} \Lambda .
\label{energy:ren}
\end{eqnarray}
By taking the limit $E \to \infty$, we find that the relation between
$\lambda_0$ and $\lambda$ has the same form in
Eq.~(\ref{lambda0:resonance}) as in the Resonance Model.
Both sides of Eq.~(\ref{energy:ren}) can be expressed 
as ratios of polynomials in $E$ of degree $N$.
The equality between these rational functions can be expressed as
$2N$ coupled nonlinear equations for the remaining $2N$ parameters
of the $N$-Resonance Model.

The amputated connected Green's function for two atoms to evolve 
into two atoms can be calculated analytically by solving the 
integral equation represented diagramatically
in Fig.~\ref{fig:inteq(RM)}.
The amplitude ${\mathcal A}$ depends only on the total energy $E$
in the center-of-mass frame:
\begin{eqnarray}
{\mathcal A} (E) = - \left[
\left( \lambda_0
    + \sum _{n=1} ^{N} \frac{ g_{0,n}^2}{E- \nu_{0,n} + i \varepsilon}
    \right)^{-1}
+ \frac{m} {4 \pi^2} \Lambda - \frac{m} {8\pi} \kappa \right]^{-1} ,
\label{Abare:N-res}
\end{eqnarray}
where $\kappa = (-mE - i \varepsilon)^{1/2}$.
Inside the square brackets in Eq.~(\ref{Abare:N-res}), the sum of
the term raised to the power $-1$ and the term $(m/4 \pi^2) \Lambda$
can be expressed as the ratio of two $N$th order polynomials in $E$.
By the fundamental theorem of arithmetic, the polynomial
in the numerator has $N$ zeroes in the complex plane,
which we denote by $\nu_n$, $n=1,...,N$.
Except on sets of the bare parameters that have measure zero,
those zeroes will be distinct.  By decomposing the inverse
of that ratio of polynomials into partial fractions,
the amplitude in Eq.~(\ref{Abare:N-res}) can be expressed
in the form
\begin{eqnarray}
{\mathcal A} (E) = - \left[
\left( \lambda
    + \sum _{n=1} ^{N}
    \frac{ g_{n}^2 }
    { E - \nu_{n} + i \varepsilon }
\right) ^{-1}
- \frac{m} {8 \pi} \kappa
\right] ^{-1} .
\label{Arenorm:N-res}
\end{eqnarray}
By comparing Eqs.~(\ref{Abare:N-res}) and (\ref{Arenorm:N-res}),
we obtain the renormalization condition in Eq.~(\ref{energy:ren}).

The condition that the Hamiltonian be hermitian requires that the 
bare parameters $\lambda_0$, $\nu_{0,n}$ and $g_{0,n}$ in the 
Hamiltonian density in Eqs.~(\ref{LH:N-resonance}) be real valued.
The expression in Eq.~(\ref{Abare:N-res}) is therefore a real-valued 
function for real $\kappa$.  This implies that the parameters 
$\nu_n$ and $g_n^2$ in Eq.~(\ref{Arenorm:N-res}) are either real
or they come in complex conjugate pairs.  This is sufficient 
to guarantee that the phase shift in Eq.~(\ref{delta0:N-resonance})
is real valued, which is necessary if the model is to describe
atoms that have no inelastic scattering channels. 
To show that the $N$-Resonance Model is renormalizable 
in the 2-atom sector for arbitrary real values 
of the parameters $\nu_n$ and $g_n$,
one must show that the renormalization
conditions in Eq.~(\ref{energy:ren}) have real solutions for 
the bare parameters $\nu_{0,n}$ and $g_{0,n}$ as $\Lambda \to \infty$.
We will verify this below for the Two-Resonance Model. 
We have not shown that this is the case for the $N$-Resonance Model 
with $N\ge 3$.

The renormalization conditions for the Two-Resonance Model 
are obtained by setting $N=2$ in Eq.~(\ref{energy:ren}).
We can use Eq.~(\ref{Z-Lambda}) to express the ultraviolet cutoff 
$\Lambda$ as a function of $\lambda$ and $Z$:
\begin{eqnarray}
\Lambda = \frac{4 \pi^2} {m \lambda}
\left( 1 - Z^{-1} \right) .
\label{LambdaZ}
\end{eqnarray}
If we also use Eq.~(\ref{lambda0:resonance}) to eliminate 
$\lambda_0$ in favor of $\lambda$ and $Z$, the renormalization 
condition in Eq.~(\ref{energy:ren}) reduces to
\begin{eqnarray}
&&
\frac{ (E - \nu_1) (E - \nu_2) }
    { (E - \nu_1) (E - \nu_2)
	+ [g_1^2 (E - \nu_2) + g_2^2 (E - \nu_1)]/\lambda }
\nonumber
\\
&& =
\frac{ (E - \nu_{0,1}) (E - \nu_{0,2})
	+ (Z^{-1}  - Z^{-2})
	[g_{0,1}^2 (E - \nu_{0,2}) + g_{0,2}^2 (E - \nu_{0,1})]/\lambda }
    { (E - \nu_{0,1}) (E - \nu_{0,2})
	+ Z^{-1} [g_{0,1}^2 (E - \nu_{0,2}) + g_{0,2}^2 (E - \nu_{0,1})]/\lambda } .
\nonumber
\\
\end{eqnarray}
We obtain four coupled equations by matching the constant terms 
and the linear terms in $E$ in both the numerator and denominator:

\begin{subequations}
\begin{eqnarray}
&& \nu_1 \nu_2 = 
\nu_{0,1} \nu_{0,2} - Z^{-1} (1 - Z^{-1})
( g_{0,1}^2 \nu_{0,2} + g_{0,2}^2 \nu_{0,1} )/\lambda ,
\\
&& \nu_1 + \nu_2 = \nu_{0,1} + \nu_{0,2}
- Z^{-1} (1 - Z^{-1} )( g_{0,1}^2 + g_{0,2}^2 )/\lambda ,
\\
&& \nu_{1} \nu_{2}
- ( g_{1}^2 \nu_{2} + g_{2}^2 \nu_{1} )/\lambda =
\nu_{0,1} \nu_{0,2}
- Z^{-1} ( g_{0,1}^2 \nu_{0,2} + g_{0,2}^2 \nu_{0,1} )/\lambda ,
\\
&& \nu_{1} + \nu_{2}
- ( g_{1}^2 + g_{2}^2 )/\lambda =
\nu_{0,1} + \nu_{0,2}
- Z^{-1} ( g_{0,1}^2 + g_{0,2}^2 )/\lambda .
\end{eqnarray}
\label{renorm:coupled}
\end{subequations}
These equations can be solved for the bare parameters 
as functions of the renormalized parameters and $Z$:
\begin{subequations}
\begin{eqnarray}
\nu_{0,1} & = &
\frac{1}{2} \left( \bar \nu_1 + \bar \nu_2 + C \right) ,
\label{nu10}
\\
\nu_{0,2} & = &
\frac{1}{2} \left( \bar \nu_1 + \bar \nu_2 - C \right) ,
\label{nu20}
\\
g_{0,1} & = & \frac{Z} {\sqrt{2}}
\left( g_{1}^2 + g_{2}^2 + \frac{D} {C}
\right) ^{1/2},
\label{g10}
\\
g_{0,2} & = & \frac{Z} {\sqrt{2}}
\left( g_{1}^2 + g_{2}^2 - \frac{D} {C}
\right) ^{1/2},
\label{g20}
\end{eqnarray}
\label{renorm:2res}
\end{subequations}
where
\begin{subequations}
\begin{eqnarray}
\bar \nu_1 & = &
\nu_{1} - (1-Z) \, g_{1}^2/\lambda ,
\\
\bar \nu_2 & = &
\nu_{2} - (1-Z) \, g_{2}^2/\lambda ,
\\
C & = &
\left[ (\bar \nu_1 - \bar \nu_2)^2
	+ 4 (1-Z)^2 \, g_{1}^2 g_{2}^2/\lambda^2 \right]^{1/2} ,
\label{C-def}
\\
D & = &
(\bar \nu_1 - \bar \nu_2)(g_{1}^2 - g_{2}^2)
- 4 (1-Z) \, g_{1}^2 g_{2}^2/\lambda .
\label{D-def}
\end{eqnarray}
\label{variables}
\end{subequations}
Note that the expressions for the bare parameters
$\nu_{0,1}$, $g_{0,1}$, $\nu_{0,2}$, and $g_{0,2}$ 
in Eq.~(\ref{renorm:2res}) are each separately
symmetric under the interchange of the renormalized parameters:
$\nu_1 \leftrightarrow \nu_2$ and $g_1 \leftrightarrow g_2$.
The Two-Resonance Model is renormalizable 
for given values of $\nu_1$, $g_1$, $\nu_2$, and $g_2$
if the solutions for 
the bare parameters in Eq.~(\ref{renorm:2res}) are real valued 
in the limit $\Lambda \to \infty$.  Since the expression for $C$ in 
Eq.~(\ref{C-def}) is the square root of a manifestly positive quantity,
it is clear that the parameters $\nu_{0,1}$ and $\nu_{0,2}$ in 
Eqs.~(\ref{nu10}) and (\ref{nu20}) are real.
The condition that the parameters $g_{0,1}$ and $g_{0,2}$ in
Eqs.~(\ref{g10}) and (\ref{g20}) are real is $g_1^2 + g_2^2 > |D|/C$.
It is easy to verify that this is satisfied for any real values of 
$\nu_1$, $\nu_2$, $g_1$, and $g_2$, even at finite $\Lambda$.

The Two-Resonance Model has been previously discussed in
Ref.~\cite{KMCWH02}, but their results for the renormalization
of the parameters of the model are incorrect.
They obtained the correct result in Eq.~(\ref{lambda0:resonance})
for the renormalization of
$\lambda$. However, their results for the renormalizations of the
parameters $\nu_{1}$ and $g_{1}$ for the first resonance
are given by the same simple equations in Eqs.~(\ref{g0:resonance})
and (\ref{nu0:resonance}) as in the Resonance Model,
while their results for the renormalizations of the parameters 
$\nu_{2}$ and $g_{2}$ for the second resonance
are given by more complicated equations.
It is easy to see from symmetry considerations
that these results cannot be correct.
The Hamiltonian of the $N$-Resonance Model is symmetric
under the interchange $\phi_1 \leftrightarrow \phi_2$
of the two molecular fields if we also interchange the bare parameters:
$\nu_{0,1} \leftrightarrow \nu_{0,2}$ and
$g_{0,1} \leftrightarrow g_{0,2}$.
Thus the expressions for the bare parameters
$\nu_{1,0}$ and $g_{1,0}$ should differ from those
for $\nu_{2,0}$ and $g_{2,0}$
simply by interchange of the renormalized parameters.
Our expressions for the renormalized parameters
in Eqs.~(\ref{renorm:2res}) satisfy this condition,
but those in Ref.~\cite{KMCWH02} do not.

The amplitude in Eq.~(\ref{Arenorm:N-res}) has poles in the energy 
variable $\kappa$ defined by Eq.~(\ref{kappa-E}).
The values of $\kappa$ at the poles are roots of the equation
\begin{eqnarray}
\left( \lambda
    - \sum _{n=1} ^{N}
    \frac{ m g_{n}^2 }
    {m \nu_{n} + \kappa^2}
\right) ^{-1}
= \frac{m} {8 \pi} \kappa.
\label{pole-mrm}
\end{eqnarray}
If the energy $E$ is real, the imaginary part of the amplitude
in Eq.~(\ref{Arenorm:N-res}) is
\begin{eqnarray}
{\rm Im} \, {\mathcal A} (E) &=&
\frac{m k}{8 \pi} \left[
\left( \lambda
    - \sum _{n=1} ^{N}
    \frac{ mg_{n}^2 }
    {m \nu_{n} - k^2} \right)^{-2}
+ \left( \frac{m k}{8 \pi} \right)^2
\right ]^{-1} \theta(E)
\nonumber \\
&+&
\frac{16 \pi^2} {m^2} \sum_i \kappa_i
\left[ 1 + \sum _{n=1} ^{N}
\frac{m^2 g_{n}^2 \kappa_i^3}{4 \pi (m \nu_{n} + \kappa_i^2)^2}
\right]^{-1}
\delta(E + \kappa_i^2/m) \, \theta(\kappa_i) ,
\nonumber
\\
\label{optical:mrm}
\end{eqnarray}
where the sum in the last term is over the 
positive real roots $\kappa_i$ of Eq.~(\ref{pole-mrm}).
This expression is positive definite, in accord with the Optical Theorem
in Eq.~(\ref{optical}). This is consistent with the absence of  
negative-norm states in the 2-atom sector of the $N$-Resonance Model.

\section*{Acknowledgments}
This research was supported in part by DOE Grants 
DE-FG02-91-ER4069, DE-FG02-05ER15715, and DE-FG02-06ER41449.

\appendix

\section{Integrals}

Some of the results in this paper require the evaluation of
integrals over the momentum $\bm{k}$ of virtual atoms of the form
\begin{eqnarray}
I_{2n}(E) = \int \frac{d^3k}{(2 \pi)^3}
\frac{k^{2n}}{E - k^2/m + i \varepsilon},
\label{In-E}
\end{eqnarray}
where $n$ is an integer.
These integrals are functions of the total energy $E$
of a pair of atoms in the center-of-mass frame.
It is convenient to express them in terms of the variable
$\kappa = (-mE - i \varepsilon)^{1/2}$
defined in Eq.~(\ref{kappa-E}).
For $n\ge 0$, the integrals given by Eq.~(\ref{In-E}) are
ultraviolet divergent.  They
can be regularized by imposing a momentum cutoff:
$|\bm{k}|< \Lambda$.
The integrals then satisfy a simple recursion relation:
\begin{eqnarray}
 I_{2n}(E) = -{m\over 2\pi^2} {\Lambda^{2n+1}\over2n+1}
-\kappa^2 \, I_{2n-2}(E).
\end{eqnarray}
This can be used to express the integrals $I_{2n}(E)$
for $n\ge 1$ in terms of $I_0(E)$, whose value is
\begin{eqnarray}
I_0(E) = - \frac{m}{2 \pi^2}
\left( \Lambda - \kappa \arctan \frac{\Lambda}{\kappa} \right).
\label{int-0}
\end{eqnarray}
The expressions for the integrals are particularly simple if we
take $\Lambda$ to be so much larger than $(m |E|)^{1/2}$ that we
can neglect terms that decrease as inverse powers of
$\Lambda/(m |E|)^{1/2}$ as $\Lambda\to \infty$.
The integral in Eq.~(\ref{int-0}) then reduces to
\begin{eqnarray}
I_0(E) = - \frac{m}{2 \pi^2}
\left( \Lambda - \frac{\pi} {2} \kappa \right).
\label{int-0:simple}
\end{eqnarray}
The next two integrals in the sequence are
\begin{subequations}
\begin{eqnarray}
I_2(E) &=& - \frac{m}{2 \pi^2}
\left( \frac{1}{3} \Lambda^3 - \Lambda \kappa^2
+ \frac{\pi} {2} \kappa^3 \right),
\label{int2}
\\
I_4(E) &=& - \frac{m}{2 \pi^2}
\left( \frac{1}{5} \Lambda^5 - \frac{1}{3} \Lambda^3 \kappa^2
    + \Lambda \kappa^4 - \frac{\pi} {2} \kappa^5 \right).
\label{int4}
\end{eqnarray}
\label{int-Lambda}
\end{subequations}

With dimensional regularization, the integral in Eq.~(\ref{In-E})
is generalized to an integral over a space with dimension
$3 - 2 \epsilon$ and then analytically continued to $\epsilon=0$.
This procedure automatically subtracts any power ultraviolet
divergences.  The resulting expressions for the integrals
can be obtained simply by setting $\Lambda = 0$ in
Eqs.~(\ref{int-0:simple}) and (\ref{int-Lambda}):
\begin{subequations}
\begin{eqnarray}
I_0(E) &=&  \frac{m}{4 \pi} \kappa,
\label{I0dr}
\\
I_2(E) &=& - \frac{m}{4 \pi} \kappa^3 ,
\\
I_4(E) &=& \frac{m}{4 \pi} \kappa^5 .
\end{eqnarray}
\label{int-dimreg}
\end{subequations}
%


\end{document}